\documentclass[12pt]{article}
\usepackage[left=1.25in, right=1.25in, top=1.25in, bottom=1.25in]{geometry}  

\usepackage{booktabs} 
\usepackage{array} 
\usepackage{paralist} 
\usepackage{verbatim} 
\usepackage{caption}
\usepackage{subcaption}
\usepackage{enumerate}
\usepackage{footmisc}
\usepackage{graphicx}
\usepackage[countmax]{subfloat}
\usepackage{mathrsfs}
\usepackage{comment}
\usepackage{url}
\usepackage[pdftex, 
   colorlinks=true, 
   linkcolor=blue, 
   citecolor=blue, 
   urlcolor=blue, 
    ]{hyperref}
\usepackage{amsmath}
\usepackage{amssymb}
\usepackage{setspace}
\usepackage{appendix}
\usepackage{natbib}
\usepackage{bm}
\usepackage{etoolbox}
\usepackage{tikz}
\usepackage{scalerel}
\usetikzlibrary{decorations.markings}
\usetikzlibrary{snakes}
\usetikzlibrary{decorations.pathreplacing,calc}
\usepackage{bm}
\usepackage{pgfplots}
\usepgfplotslibrary{fillbetween}
\usetikzlibrary{backgrounds,matrix,fit}
\usetikzlibrary{fadings,hobby,calc}
\usetikzlibrary{decorations.pathreplacing}
\tikzset{%
  show curve controls/.style={
    postaction={
      decoration={
        show path construction,
        curveto code={
          \draw [blue]
            (\tikzinputsegmentfirst) -- (\tikzinputsegmentsupporta)
            (\tikzinputsegmentlast) -- (\tikzinputsegmentsupportb);
          \fill [red, opacity=0.5]
            (\tikzinputsegmentsupporta) circle [radius=.5ex]
            (\tikzinputsegmentsupportb) circle [radius=.5ex];
        }
      },
      decorate
}}}

\usepackage{amsmath,fouriernc}

\pgfplotsset{compat=1.15}
\usetikzlibrary{arrows}
\definecolor{rvwvcq}{rgb}{0.08235294117647059,0.396078431372549,0.7529411764705882}
\definecolor{ttqqqq}{rgb}{0.2,0,0}
\definecolor{ffqqqq}{rgb}{1,0,0}
\definecolor{qqqqff}{rgb}{0,0,1}
\definecolor{ffzzqq}{rgb}{1,0.6,0}
\definecolor{sexdts}{rgb}{0.1803921568627451,0.49019607843137253,0.19607843137254902}
\definecolor{qqttcc}{rgb}{0,0.2,0.8}
\definecolor{ududff}{rgb}{0.30196078431372547,0.30196078431372547,1}
\definecolor{xdxdff}{rgb}{0.49019607843137253,0.49019607843137253,1}
\definecolor{zzttqq}{rgb}{0.6,0.2,0}

\makeatletter
\renewcommand{\section}{\@startsection{section}{1}{0mm}{-1.5\baselineskip}{0.8\baselineskip}{\normalfont\large\centering}}
\renewcommand{\subsection}{\@startsection{subsection}{2}{0mm}{-0.1\baselineskip}{0.5\baselineskip}{\normalfont\bf\flushleft}}
\renewcommand{\@seccntformat}[1]{\csname the#1\endcsname \hspace{+0mm}\large{.}\hspace{+1.9mm}}
\renewcommand{\@seccntformat}[2]{\csname the#1\endcsname \hspace{+0mm}\large{.}\hspace{+1.9mm}}
\makeatother

\renewcommand{\theequation}{\arabic{equation}}

\newlength{\extraspace}
\newlength{\extraspaces}
\newcounter{dummy}

\newcommand{\baa}{
\addtocounter{equation}{1} \setcounter{dummy}{\value{equation}}
\setcounter{equation}{0}
\renewcommand{\theequation}{\arabic{dummy}\alph{equation}}
\begin{eqnarray}
\addtolength{\abovedisplayskip}{\extraspaces}
\addtolength{\belowdisplayskip}{\extraspaces}
\addtolength{\abovedisplayshortskip}{\extraspace}
\addtolength{\belowdisplayshortskip}{\extraspace}}

\newcommand{\eaa}{
\end{eqnarray}
\setcounter{equation}{\value{dummy}}
\renewcommand{\theequation}{\arabic{equation}}}

\newcommand{\be}{\begin{equation}
\addtolength{\abovedisplayskip}{\extraspaces}
\addtolength{\belowdisplayskip}{\extraspaces}
\addtolength{\abovedisplayshortskip}{\extraspace}
\addtolength{\belowdisplayshortskip}{\extraspace}}
\newcommand{\ee}{\end{equation}}

\newcommand{\ba}{\begin{eqnarray}
\addtolength{\abovedisplayskip}{\extraspaces}
\addtolength{\belowdisplayskip}{\extraspaces}
\addtolength{\abovedisplayshortskip}{\extraspace}
\addtolength{\belowdisplayshortskip}{\extraspace}}
\newcommand{\ea}{\end{eqnarray}}

\newcommand{\bd}{\begin{displaymath}
\addtolength{\abovedisplayskip}{\extraspaces}
\addtolength{\belowdisplayskip}{\extraspaces}
\addtolength{\abovedisplayshortskip}{\extraspace}
\addtolength{\belowdisplayshortskip}{\extraspace}}
\newcommand{\ed}{\end{displaymath}}

\newcommand{\argmin}[1]{\underset{#1}{\operatorname{arg}\!\operatorname{min}}\;}
\newcommand{\argmax}[1]{\underset{#1}{\operatorname{arg}\!\operatorname{max}}\;}
\newcommand{\deel}[2]{{\textstyle{#1 \over #2}}}
\newcommand{\hf}{{\textstyle{1\over 2}}}

\makeatletter

\patchcmd{\NAT@citex}
  {\@citea\NAT@hyper@{%
     \NAT@nmfmt{\NAT@nm}%
     \hyper@natlinkbreak{\NAT@aysep\NAT@spacechar}{\@citeb\@extra@b@citeb}%
     \NAT@date}}
  {\@citea\NAT@nmfmt{\NAT@nm}%
   \NAT@aysep\NAT@spacechar\NAT@hyper@{\NAT@date}}{}{}

\patchcmd{\NAT@citex}
  {\@citea\NAT@hyper@{%
     \NAT@nmfmt{\NAT@nm}%
     \hyper@natlinkbreak{\NAT@spacechar\NAT@@open\if*#1*\else#1\NAT@spacechar\fi}%
       {\@citeb\@extra@b@citeb}%
     \NAT@date}}
  {\@citea\NAT@nmfmt{\NAT@nm}%
   \NAT@spacechar\NAT@@open\if*#1*\else#1\NAT@spacechar\fi\NAT@hyper@{\NAT@date}}
  {}{}
\makeatother

\newtheorem{theorem}{Theorem}
\newtheorem{lemma}{Lemma}

\newtheorem{definition}{Definition}

\newtheorem{corollary}{Corollary}

\newtheorem{example}{Example}

\newcommand{\q}[1]{``#1''}

\begin{document}

\title{{\LARGE\sc Multivariate Majorization}\\[-1mm]{\LARGE\sc in Principal--Agents Models}\\[6mm]}
\author{\large Nicholas C. Bedard, Jacob K. Goeree, and Ningyi Sun\thanks{Bedard: Wilfrid Laurier University
75 University Avenue W., Waterloo, Ontario, Canada; Email: \href{mailto:nbedard@wlu.ca}{nbedard@wlu.ca}. Goeree: AGORA Center for Market Design, UNSW Australia Business School, Sydney NSW 2052, Australia; Email: \href{mailto: jacob.goeree@gmail.com}{jacob.goeree@gmail.com}. Sun: Economics Department, Business School, Monash University; Email: \href{mailto: ningyi.sun.econ@gmail.com}{ningyi.sun.econ@gmail.com}. Goeree gratefully acknowledges funding from the Australian Research Council (DP190103888 and DP220102893). We thank seminar audiences at the 2016 APIOS in Melbourne, the University of Western Ontario, the University of Guelph and the University of Waterloo for insightful comments and discussions.}}
\date{\small\today}
\maketitle
\vspace*{5mm}

\begin{abstract}
\begin{singlespace}
\noindent We introduce a definition of multivariate majorization that is new to the economics literature.  Our majorization technique allows us to generalize Mussa and Rosen's (1978) ``ironing'' to a broad class of multivariate principal-agents problems. Specifically, we consider adverse selection problems in which agents' types are one dimensional but informational externalities create a multidimensional ironing problem. Our majorization technique applies to discrete and continuous type spaces alike and we demonstrate its usefulness for contract theory and mechanism design. We further show that multivariate majorization yields a natural extension of second-order stochastic dominance to multiple dimensions and derive its implications for decision making under multivariate risk.
\end{singlespace}
\end{abstract}

\vfill
\thispagestyle{empty}
\newpage
\addtocounter{page}{-1}
\addtolength{\baselineskip}{1.4mm}

\section{Introduction}

We consider principal-agent(s) models with adverse selection and informational externalities. Examples include the design of mass-produced goods with interdependent values, the production of public or club goods, and multi-agent contracting. In each of these cases, information held by one agent may be pertinent to another. For instance, when a next generation smart phone comes out, one buyer may know about its processing speed while another knows about its screen's resolution. Both attributes determine the value of the product for each buyer, possibly in different ways as buyers need not agree on what makes a product valuable.

The principal is not privy to agents' private information and must elicit this information to make an optimal decision. Agents optimally reveal their private information truthfully if the principal's decision rule satisfies certain ``incentive compatibility'' constraints, which can typically be reduced to monotonicity constraints on the allocation rule. Moreover, with \textit{ex post} incentive constraints the principal's optimal decision is robust in that it does not rely on agents' beliefs.\footnote{Under Bayesian incentive constraints, monotonicity is only required on the interim expected allocation rules. Nevertheless, the system of monotonicity constraints still needs to be solved as a whole and the problem therefore remains multidimensional. Our proposed technique, with some modification, would still be needed.} The principal's problem for a broad class of applications can thus be formulated as
\be \label{problem: original problem}
\max_{\substack{\rule{0pt}{10pt}q\,:\,\mathbf{X}\,\rightarrow\,\mathbb{R}_{\geq 0}\\\rule{0pt}{8pt}\text{s.t. $q(\mathbf{x})$ is non-decreasing}\\\rule{0pt}{8pt}\text{in each coordinate}}}
\hspace*{-5mm}\mathbb{E}\bigl[q(\mathbf{x})\alpha(\mathbf{x})-C(q(\mathbf{x}))\bigr]
\ee
where the multidimensional type set $\mathbf{X}=\prod_{i=1}^N X_i$ is a finite subset (or compact subset) of $\mathbb{R}^N$, $\alpha$ is an exogenously specified continuous function and $C$ is a convex function of the principal's decision variable $q$. The monotonicity condition that $q$ be non-decreasing in each coordinate reflects the requirement of incentive compatible (or equilibrium) behavior by the agents.

In the ``regular'' case, i.e. when $\alpha$ is non-decreasing in each coordinate, the solution to \eqref{problem: original problem} simply follows from first-order conditions. However, when informational externalities exist, the regularity assumption is overly restrictive and in many important applications $\alpha$ is not a non-decreasing function. This
raises the question how to solve the principal's problem in \eqref{problem: original problem}.

For a \textit{one}-dimensional model of quality choice, \cite{MussaRosen1978} first showed that the solution to the principal's problem follows by replacing $\alpha$ with an ``ironed'' version that is non-decreasing so that standard first-order conditions apply. Mussa and Rosen's ironing technique has found ubiquitous application in economics and finance.\footnote{Mussa and Rosen's (\citeyear{MussaRosen1978}) paper has almost 4,500 Google Scholar citations.} However, their methodology cannot be immediately applied to a setting with informational externalities, which is naturally multidimensional.\footnote{\cite{RoughgardenTalgam-Cohen2016}, for example, note that a naive generalization of ironing in a model with interdependent values fails to properly account for incentive-compatibility constraints.} Likewise, the ironing technique proposed by \cite{Myerson1981}, which entails convexifying the primitive of $\alpha$, does not readily extend to multidimensional settings.

In this paper we develop a constructive approach to multidimensional ironing. We extend Mussa and Rosen's (\citeyear{MussaRosen1978}) ironing approach by manipulating the $\alpha$ function in \eqref{problem: original problem} to generate an alternative problem whose unconstrained solution is the same as the solution to the original constrained problem.
To accomplish this, we reinterpret multidimensional ironing in terms of multivariate majorization. This generalizes insights developed by \cite{GoereeKushnir2022} who first noted the connection between one-dimensional ironing and univariate majorization. While the result is not a closed-form solution to the original problem \eqref{problem: original problem}, we do provide a convenient and easily implementable algorithm for calculating the solution as well as some its general characteristics.

Several definitions of multivariate majorization exist in the mathematics literature, e.g. row-majorization, column-majorization, and linear-combinations majorization \citep[see][Chapter 15]{MarshallOlkinArnold2010}, but none of these are relevant for solving the principal's problem in \eqref{problem: original problem}. We exploit a notion of multivariate majorization that generalizes univariate majorization to higher dimensions.\footnote{See also \cite{Hwang1979} and \cite{BrualdiDahl2013} who apply this notion to arbitrary partially-ordered sets.} Recall that if $g:X\rightarrow\mathbb{R}$ and $h:X\rightarrow\mathbb{R}$ are non-decreasing functions over some finite ordered type set  $X=\{x^1,\ldots,x^K\}$ then $h$ majorizes $g$, denoted $h\succ g$, iff \begin{equation}\label{univariateMajorization}
  \sum_{j\,=\,1}^kg(x^j)\,\geq\,\sum_{j\,=\,1}^kh(x^j)
\end{equation}
for $k=1,\ldots,K$ with equality when $k=K$. The relevance of majorization for solving the principal's problem in the univariate case stems from two classic results by \cite{HardyLittlewoodPolya1929}. The first is that $h\succ g$ iff $g(x)=\sum_{x'\in X}T(x,x')h(x')$ where $T$ is a doubly-stochastic operator. The second is that $h\succ g$ iff
\begin{equation}\label{schurConvex}
  \sum_{x\,\in\,X}\phi(g(x))\,\leq\,\sum_{x\,\in\,X}\phi(h(x))
\end{equation}
for any convex function $\phi:\mathbb{R}\rightarrow\mathbb{R}$ where the inequality is strict if $\phi$ is strictly convex and $g\neq h$. \cite{GoereeKushnir2022} exploit both properties to show that for a given function $\alpha$, not necessarily non-decreasing, there exists a unique $\overline{\alpha}\succ\alpha$ that is the minimum with respect to majorization order, i.e. for any other $\alpha'\succ\alpha$ we have $\alpha'\succ\overline{\alpha}$.

Our definition of multivariate majorization generalizes the lower sums in \eqref{univariateMajorization} to sums over \textit{lower subsets} of the multidimensional type set $\mathbf{X}$ in \eqref{problem: original problem}. For non-decreasing functions $g:\mathbf{X}\rightarrow\mathbb{R}$ and $h:\mathbf{X}\rightarrow\mathbb{R}$ we show that $h\succ g$ iff $g(\mathbf{x})=\sum_{\mathbf{x}'\in\mathbf{X}}T(\mathbf{x},\mathbf{x}')h(\mathbf{x}')$ where $T$ belongs to a restricted class of (orthogonal) doubly stochastic operators. Unlike the univariate case, for a given multivariate function $\alpha$ there does not necessarily exist a unique $\overline{\alpha}\succ\alpha$ that is the minimum with respect to the majorization order, i.e. there may exist $\alpha'\succ\alpha$ for which neither $\alpha'\succ\overline{\alpha}$ nor $\overline{\alpha}\succ\alpha'$. We show, however, that the $\overline{\alpha}$ that is relevant for solving \eqref{problem: original problem} is a \textit{minimal element} with respect to the majorization order, i.e. if there exists $\alpha'$ such that $\overline{\alpha}\succ\alpha'\succ\alpha$ then
$\alpha'=\overline{\alpha}$. We exploit a generalization of \eqref{schurConvex} to the multivariate case to select a unique function, $\overline{\alpha}$, from the set of minimal elements. The principal's optimal decision when faced with $\overline{\alpha}$ then readily follows from first-order conditions and solves the original problem in \eqref{problem: original problem}.

\subsection{Related Literature}

One of the earliest attempts to deal with multidimensional incentive-compatibility constraints in a general setting is \cite{RochetChone1998}. In their environment, a monopolist with strictly convex costs sells to a consumer with multidimensional preferences, each dimension corresponding to an attribute of the good designed by the seller. They introduce the concept of ``sweeping.'' Roughly, sweeping is an operation that redistributes the density of a measure in a way that preserves the original center-of-mass of the measure. \cite{RochetChone1998} do not use sweeping to construct a solution but instead its use is to verify whether a candidate incentive-compatible solution is optimal. Specifically, their results say that if (the derivative of) the first-order condition of the seller's unconstrained problem, evaluated at the candidate solution, is non-zero but can be ``swept'' to zero then the candidate solution is optimal. In a similar environment, \cite{DDT2017} deal with multidimensional incentive-compatibility constraints through an optimal transport problem that is dual to the mechanism design problem. Their analysis shows that accommodating the incentive compatibility constraints requires the application of mean-preserving spreads on measures in the dual problem, echoing \cite{RochetChone1998}'s sweeping results.

Some recent progress has been made on multidimensional ironing in multidimensional monopolist problem environments, in particular by \cite{Carroll2017}, \cite{CDW2019} and \cite{HaghpanahHartline2021}. These papers construct generalized virtual values based on the Lagrangian dual of the monopolist's problem. \cite{Carroll2017} is interested in the mechanism that maximizes the monopolist's worst-case expected profits over all possible joint beliefs consistent with known marginal beliefs. \cite{CDW2019} seek simple mechanisms that are approximately optimal. \cite{HaghpanahHartline2021} characterize when pure bundling is optimal for the monopolist.

Our paper differs from these approaches in that agents' types are one dimensional but informational externalities create multidimensional incentive constraints. Another difference is that we develop a constructive, and computationally simple, method for multidimensional ironing.

\subsection{Organization}

The next section introduces our novel definition of multivariate majorization (Section 2.1) and shows how it can be used to implement ironing in the multidimensional case (Section 2.2). We also show how to extend the results to continuous type spaces (Section 2.3). Section 3 introduces access rights that allow the principal to screen out some types, e.g. to increase revenues. We show how our multivariate majorization approach can be extended to include probabilistic or all-or-nothing access rights. Section 4 discusses applications to decision making under multivariate risk, the optimal design of mass-produced goods, as well as multi-agent contracting. Section 5 concludes. Proofs can be found in Appendix A.

\section{Multivariate Majorization and Ironing}

Let $\mathcal{N}=\{1,\ldots,N\}$ denote the set of agents. It is instructive to start with the discrete case, i.e., $\mathbf{X}=\prod_{i=1}^N X_i$ where each $X_i$ is an ordered finite set. For $i\in\mathcal{N}$, the type $x_i$ is drawn from $X_i$ according to distribution $F_i(x_i)$ with probability function (or density if $X_i$ is continuous) $f_i(x_i)$. We make no assumptions about $X_i$ or $F_i$ but assume independence across bidders, i.e. $\mathbf{x}$ is drawn according to $F(\mathbf{x})=\prod_{i\in N}F_i(x_i)$. For a profile of types we write $\mathbf{x}=\{x_i,\mathbf{x}_{-i}\}$ with $\mathbf{x}_{-i}=\{x_1,\ldots,x_{i-1},x_{i+1},\ldots, x_n\}$, and $\mathbf{X}_{-i}=\prod_{j\in N\setminus\{i\}}X_j$.  For $\mathbf{x},\mathbf{y}\in\mathbf{X}$, we write $\mathbf{x}\leq\mathbf{y}$ ($\mathbf{x}<\mathbf{y}$) if $x_i \leq y_i$ ($x_i < y_i$) for all $i\in\mathcal{N}$ and define $[\mathbf{x},\mathbf{y}]=\{\mathbf{e}\in X\mid\mathbf{x}\,\leq\,\mathbf{e}\,\leq\,\mathbf{y}\}$.

Let $\underline x_i=\min X_i$ and $\overline{x}_i= \max X_i$. For an arbitrary function $g:\mathbf{X}\rightarrow\mathbb{R}$ we define the partial derivatives $\underline{\Delta}_ig(\mathbf{x})=g(x_i,\mathbf{x}_{-i})-g(x^-_i,\mathbf{x}_{-i})$ and $\overline{\Delta}_ig(\mathbf{x})=g(x^+_i,\mathbf{x}_{-i})-g(x_i,\mathbf{x}_{-i})$ where $x^+_i$ $(x^-_i)$ is the type just above (below) $x_i$ with the convention that $\underline{\Delta}_ig(\underline{x}_i,\mathbf{x}_{-i})=g(\underline{x}_i,\mathbf{x}_{-i})$ and $\overline{\Delta}_ig(\overline{x}_i,\mathbf{x}_{-i})=0$. (In the case of continuous types, $\underline{\Delta}_ig(\mathbf{x})=\overline{\Delta}_ig(\mathbf{x})=\partial_{x_i}g(\mathbf{x})$.) A function $g:\mathbf{X}\rightarrow\mathbb{R}$ is non-decreasing if $\mathbf{x}\leq\mathbf{y}$ implies $g(\mathbf{x})\leq g(\mathbf{y})$. Let $\mathbb{E}[g(\mathbf{x})]=\sum_{\mathbf{x}\in\mathbf{X}}f(\mathbf{x})g(\mathbf{x})$.

\subsection{Multivariate Majorization}

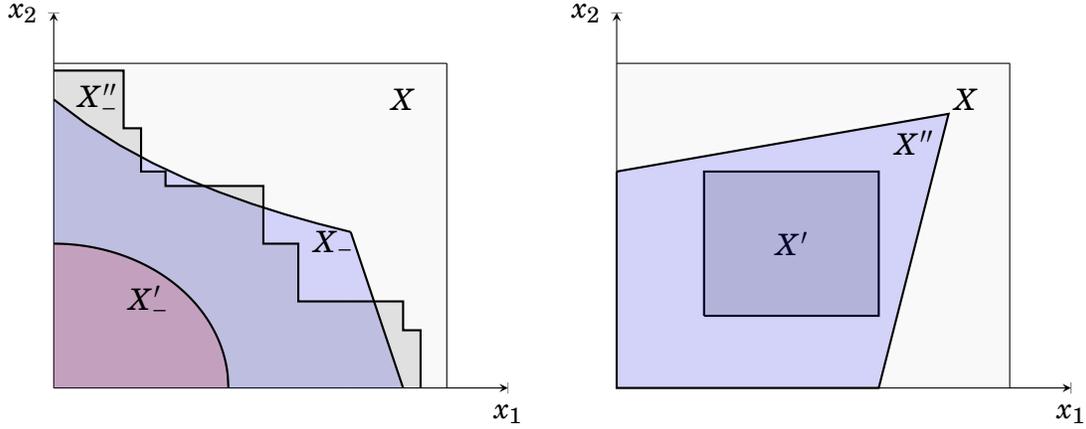
\begin{figure}[t]
 \centering
\begin{tikzpicture}
  \begin{axis}[
  width=0.5\textwidth,
  axis lines = left,
  xmax=1.3, ymax=1.3,
  xmin=0, ymin=0,
  xtick={1.3}, xticklabels={$x_1$}, xlabel={},
  ytick={1.3},yticklabels={$x_2$}, ylabel={},
  samples=10,
  clip=false,
  scatter/classes={%
    a={draw=black}},
  ]
 \path[name path=xaxis] (axis cs:0,0) -- (axis cs:1,0);
 \draw[name path=a] (axis cs:0,1.125) -- (axis cs:1.125,1.125);
 \path[name path=b] (axis cs:0,0) -- (axis cs:1.125,0);
 \addplot[fill=gray, fill opacity=0.05, domain=0:1.125] fill between[
        of=a and b,
        soft clip={domain=0:1.125},
    ];
 \draw (axis cs:1.125,1.125)--(axis cs:1.125,0);
 \path (axis cs:1,1)  node  [black]{$X$};

  \path[name path=xaxis2] (axis cs:0.85,0) -- (axis cs:1,0);

 \addplot[name path=f1, domain=0:0.85, thick, black]{1/(x+1)};
 \addplot[name path=f2, domain=0.85:1, thick, black]{-x*((1/(1.85)))/(0.15)+0.85*((1/(1.85)))/(0.15)+(1/(1.85))};
 \addplot[name path=yaxis, domain=0:0.85, blue,opacity=0]{x*(1/(1.85))/0.85};
 \addplot[
        color=blue,
        fill=blue,
        fill opacity=0.15,
        domain=0:0.85
    ]
    fill between[
        of=xaxis and f1,
        soft clip={domain=0:0.85},
    ];
     \addplot [
        color=blue,
        fill=blue,
        fill opacity=0.15,
        domain=0.85:1
    ]
    fill between[
        of=xaxis2 and f2,
        soft clip={domain=0.85:1},
    ];
 \path (axis cs:0.8,0.5)  node  [black]{$X_-$};

    \path[name path=xaxis3] (axis cs:0,0) -- (axis cs:0.5,0);
    \addplot[name path=f3, domain=0:0.5, thick, black, samples=250]{(0.25-x*x)^(0.5)};
    \addplot[
    		color=red,
		fill=red,
		fill opacity=0.15,
		domain=0:0.5
		]
	fill between [
		of=xaxis3 and f3,
		soft clip={domain=0:0.5},
		];
    \path (axis cs:.27,0.3) node [black]{$X_-'$};

    \addplot[name path=step, scatter,%
    scatter src=explicit symbolic, black, thick, no marks]%
table[meta=label] {
x	 y		 	label
0		1.1		a
0.2 		1.1		a
0.2		0.9		a
0.25		0.9		a
0.25		0.75 		a
0.32		0.75 		a
0.32		0.7 		a
0.6		0.7 		a
0.6		0.5 		a
0.7		0.5 		a
0.7		0.3 		a
1		0.3 		a
1		0.2 		a
1.05		0.2 		a
1.05		0		a
    };
   \addplot[fill=black, fill opacity=0.1, domain=0:1.125] fill between[
        of=xaxis and step,
    ];
   \path (axis cs:.1/.8,1) node [black]{$X_-''$};
    \end{axis}
    \end{tikzpicture}
    \hfil%
\begin{tikzpicture}
	\begin{axis}[
 		width=0.5\textwidth,
		axis lines = left,
		xmax=1.3, ymax=1.3,
		xmin=0, ymin=0,
		xtick={1.3}, xticklabels={$x_1$}, xlabel={},
		ytick={1.3},yticklabels={$x_2$}, ylabel={},
		samples=10,
		clip=false,
  		scatter/classes={%
    		a={draw=black}},
 		]
	\path[name path=xaxis] (axis cs:0,0) -- (axis cs:1,0);
	\draw[name path=a] (axis cs:0,1.125) -- (axis cs:1.125,1.125);
	\path[name path=b] (axis cs:0,0) -- (axis cs:1.125,0);
	\addplot[fill=gray, fill opacity=0.05, domain=0:1.125] fill between[
	of=a and b,
	soft clip={domain=0:1.125},
	];
	\draw (axis cs:1.125,1.125)--(axis cs:1.125,0);
	\path (axis cs:1,1)  node  [black]{$X$};

	\path[name path=xaxis2] (axis cs:0.85,0) -- (axis cs:1,0);
	
	\draw[fill=black, fill opacity=0.15, thick] (axis cs:0.25,0.25) -- (axis cs:.25,0.75)--(axis cs:.75,0.75)--(axis 			cs:.75,0.25)--(axis cs:.25,0.25);
	\path (axis cs:.5,.5)  node  [black]{$X'$};

	\draw[fill=blue, fill opacity=0.15, thick] (axis cs:0,0.75) -- (axis cs:0.95,0.95)--(axis cs:.75,0)--(axis cs:0,0)--(axis cs:0,0.75);
	\path (axis cs:.85,.85)  node  [black]{$X''$};
	\end{axis}
    \end{tikzpicture}
\vspace*{-2mm}
\caption{In the left panel $\mathbf{X}_-$, $\mathbf{X}'_-$ and $\mathbf{X}_-''$ are all lower sets of $\mathbf{X}$. In the right panel, $\mathbf{X}'$ and $\mathbf{X}''$ are not lower sets of $\mathbf{X}$.} \label{uppersets}
\end{figure}

We start with the multivariate version of a lower set.
\begin{definition}
	A subset $\mathbf{X}_-\subseteq\mathbf{X}$ is a \textbf{lower set} of $\mathbf{X}$ if $\mathbf{x}\in\mathbf{X}_-$ and $\mathbf{y}\in\mathbf{X}$ with $\mathbf{y}\leq\mathbf{x}$ implies $\mathbf{y}\in\mathbf{X}_-$.
\end{definition}
For instance, each of the subsets in the left panel of Figure \ref{uppersets} is an example of a lower set, while neither subset in the right panel is.
\begin{definition}\label{multiVariate}
For $g:\mathbf{X}\rightarrow\mathbb{R}$ and $h:\mathbf{X}\rightarrow\mathbb{R}$, $h$ \textbf{majorizes} $g$, denoted $h\succ g$, if for any lower set $\mathbf{X}_-\subset\mathbf{X}$ we have
$\mathbb{E}[g(\mathbf{x})|\mathbf{x}\in\mathbf{X}_-]\geq\mathbb{E}[h(\mathbf{x})|\mathbf{x}\in\mathbf{X}_-]$ and $\mathbb{E}[g(\mathbf{x})]=\mathbb{E}[h(\mathbf{x})]$.
\end{definition}
This definition is not restricted to non-decreasing functions as this would make it irrelevant for developing a multidimensional version of Mussa and Rosen's (\citeyear{MussaRosen1978}) ironing method.\footnote{Definition \ref{multiVariate} can be motivated by a Kuhn-Tucker approach, see an earlier version of this paper \citep{BGS2020}. While it is not surprising that non-decreasingness constraints can be captured by a Kuhn-Tucker program, it is surprising that the resulting conditions form a well-defined mathematical structure that generalizes univariate majorization. The multivariate majorization notion of Definition \ref{multiVariate} offers a conceptual and computational approach to multidimensional ironing.}

It will be useful below to paramterize the majorization relation between two functions. To do so we will adapt a well-known concept in the univariate majorization literature call a $T$-transform: a doubly stochastic operator of the form $T=\alpha I+(1-\alpha)P$ with $I$ the identity, $P$ a permutation that interchanges only two elements, and $0\leq\alpha\leq 1$. Under a $T$-transform the sequence $(g(x^1),\ldots,g(x^K))$ maps to $(g(x^1),\ldots,\alpha g(x^i)+(1-\alpha)g(x^j),\ldots,\alpha g(x^j)+(1-\alpha)g(x^i),\ldots,g(x^K))$, i.e. there is a transfer $(1-\alpha)(g(x^j)-g(x^i))$ from type $j$ to $i$.\footnote{If $g$ represents \q{wealth} and is non-decreasing in type then this amounts to a non-negative transfer from a \q{wealthier} to a \q{poorer} type (i.e. it is a \q{Robin Hood} transfer).}

\cite{Muirhead1903} showed that for non-decreasing univariate functions $g$ and $h$, a necessary and sufficient condition for $h\succ g$ is that $g$ can be obtained from $h$ via a series of $T$-transforms. We obtain a similar result below for functions defined over the multidimensional set $\mathbf{X}$ if we restrict transfers to be between $(x_i,{\bf x}_{-i})$ and $(x'_i,{\bf x}_{-i})$ for some $i\in N$, $x_i,x'_i\in X_i$, and ${\bf x}_{-i}\in\mathbf{X}_{-i}$. We refer to $T$-transforms of this type as \textit{orthogonal} $T$-transforms.
\begin{theorem} \label{iffTtrans}
Let $g(\mathbf{x})$ and $h(\mathbf{x})$ be non-decreasing in each coordinate. Then $h\succ g$ if and only if $g(\mathbf{x})$ can be obtained from $h(\mathbf{x})$ via a series of orthogonal $T$-transforms.
\end{theorem}
For the univariate case, \cite{HardyLittlewoodPolya1929} sharpened Muirhead's (\citeyear{Muirhead1903}) result to $g\succ h$ if and only if $g(x)=\sum_{x'\in X}T(x,x')h(x')$ for some doubly stochastic operator $T$. But in the multivariate case, the orthogonality requirement restricts the possible doubly-stochastic transformations. In particular, orthogonality does not allow the transformation to mix between arbitrary elements of $X$. To illustrate, suppose $N=2$ and $\mathbf{X}=\{\underline{x},\overline{x}\}\times\{\underline{x},\overline{x}\}$ so we can represent $g:\mathbf{X}\rightarrow\mathbb{R}$ by a $2\times 2$ matrix where the rows correspond to agent 1's types and the columns to agent 2's types. If, for instance,
$g=\bigl(\!\begin{array}{cc} 0 \!&\! 2 \\[-0.5mm] 4 \!&\! 6 \end{array}\!\bigr)$ and $h$ is obtained from $g$ by averaging along columns then $g\succ h$ but not if we average along the diagonal, i.e. $\bigl(\!\begin{array}{cc} 0 \!&\! 2 \\[-0.5mm] 4 \!&\! 6 \end{array}\!\bigr)\succ \bigl(\!\begin{array}{cc} 1 \!&\! 1 \\[-0.5mm] 5 \!&\! 5 \end{array}\!\bigr)$ but
$\bigl(\!\begin{array}{cc} 0 \!&\! 2 \\[-0.5mm] 4 \!&\! 6 \end{array}\!\bigr)\not\succ \bigl(\!\begin{array}{cc} 0 \!&\! 3 \\[-0.5mm] 3 \!&\! 6 \end{array}\!\bigr)$.

The multivariate majorization order $\succ$ in Definition \ref{multiVariate} is a preorder. Given $\alpha:\mathbf{X}\rightarrow\mathbb{R}$, we say that a non-decreasing function $\overline{\alpha}:\mathbf{X}\rightarrow\mathbb{R}$ is a \textit{minimal element} that majorizes $\alpha$ if the existence of a non-decreasing function $g:\mathbf{X}\rightarrow\mathbb{R}$ such that $\overline{\alpha}\succ g \succ \alpha$ implies $g=\overline{\alpha}$. As we show below, the properly ironed $\alpha$ will be a minimal element that majorizes $\alpha$. In the univariate case, this minimal element is unique  \citep[see][]{GoereeKushnir2022}. In the multivariate case, there may exist multiple minimal elements, as we show in the following example.

\begin{example}[Minimal Elements]\label{lambdamd}
{\em Suppose $X_1=X_2=\{x_1,x_2\}$ and types are uniformly distributed. Let
\begin{displaymath}
\alpha\,=\,\biggl(\!\begin{array}{cc} 6 & 0 \\ 0 & 6 \end{array}\!\biggr)
\end{displaymath}
where the rows and columns correspond to $X_1$ and $X_2$ respectively. It is straightforward to verify that
\begin{displaymath}
	g \,=\, \biggl(\!\begin{array}{cc}
	1 & 2 \\ 2 & 7
	\end{array}\!\biggr)
\end{displaymath}
is a non-decreasing function that majorizes $\alpha$. It is, however, not minimal. To find all minimal elements we parameterize
\begin{displaymath}
h\,=\,\biggl(\!\begin{array}{cc} 6-\lambda-\mu & \mu \\ \lambda & 6 \end{array}\!\biggr)
\end{displaymath}
with $\lambda,\mu\geq 0$ so that $h\succ\alpha$ (i.e. a transfer from each agent's lowest type to her highest type). The lower boundary of the set determined by the constraints that ensure $h$ is non-decreasing, i.e. $6-\lambda-\mu\leq\min(\lambda,\mu)$ and $\max(\lambda,\mu)\leq 6$, corresponds to the set of minimal elements, which form the polyline $\mu=\max(6-2\lambda,3-\hf\lambda)$ for $\lambda\in[0,6]$.} Below we determine how to choose the correct minimal element as the ironed version of $\alpha$. $\hfill\blacksquare$
\end{example}

\subsection{Multivariate Ironing}

In this section, we determine the ironed $\overline{\alpha}$ such that the relaxed problem
\be \label{problem: relaxed problem}
\max_{\rule{0pt}{10pt}q\,:\,\mathbf{X}\,\rightarrow\,\mathbb{R}_{\geq 0}}
\hspace*{0mm}\mathbb{E}\bigl[q(\mathbf{x})\overline{\alpha}(\mathbf{x})-C(q(\mathbf{x}))\bigr]
\ee
is equivalent to \eqref{problem: original problem}, i.e. it has the same solution and optimal value. We show that $\overline{\alpha}$ can be obtained via a convex minimization problem that selects a unique member from the set of minimal elements. Uniqueness of $\overline{\alpha}$ then implies the solution, $q^*(\mathbf{x})$, to \eqref{problem: relaxed problem} is also unique.

Theorem \ref{form} below establishes the existence of a function $\overline{\alpha}$ that is non-decreasing and is a minimal element that majorizes $\alpha$. We also describe how $\overline{\alpha}$ generates a particular partition of $X$. We write $[\mathbf{x},\mathbf{y}]\in\mathbf{S}$ if $\mathbf{x}\leq\mathbf{e}\leq\mathbf{y}$ implies $\mathbf{e}\in\mathbf{S}$.
\begin{definition}
	$\mathbf{S}\subseteq\mathbf{X}$ is ultramodular if $\mathbf{x},\mathbf{y}\in\mathbf{S}$ and $\mathbf{x}\leq\mathbf{y}$ implies $[\mathbf{x},\mathbf{y}]\in\mathbf{S}$. A partition $\mathscr{P}$ of $\mathbf{X}$ is ultramodular if all its cells are ultramodular.
\end{definition}
Intuitively, $\mathbf{S}\subseteq\mathbf{X}$ is ultramodular if for any line parallel to one of the axes or of positive slope, the intersection with $\mathbf{S}$ is empty, a point, or a single segment. (Akin to the definition of convex sets for which the intersection with \textit{any} line is empty, a point, or a single segment.)
\begin{theorem}\label{form}
The solution to \eqref{problem: original problem} is given by
\be
q^*(\mathbf{x})\,=\,C'^{-1}\bigl(\max\{\overline{\alpha}(\mathbf{x}),0\}\bigr)
\ee
where $\overline{\alpha}\succ\alpha$ is the unique minimal element that solves
\be \label{eq: ironed alpha}
\overline{\alpha}\,=\,\argmin{g\,:\,\mathbf{X}\,\rightarrow\,\mathbb{R} \atop g\,\succ\,\alpha}\mathbb{E}\bigl[\phi(g(\mathbf{x}))\bigr]
\ee
for any strictly convex function $\phi:\mathbb{R}\rightarrow\mathbb{R}$ with $\phi(0)=\phi'(0)=0$.

The solution $\overline{\alpha}$ is non-decreasing and its level sets form an ultramodular partition $\mathscr{P}$ of $\mathbf{X}$: for $\mathbf{x}\in\mathbf{X}$, the cell of $\mathscr{P}$ containing $\mathbf{x}$ is $P(\mathbf{x})=\{\mathbf{y}\,|\,\overline{\alpha}(\mathbf{y})=\overline{\alpha}(\mathbf{x})\}$ and $\overline{\alpha}(\mathbf{x})=\mathbb{E}[\alpha(\mathbf{y})\,|\,\mathbf{y}\in P(\mathbf{x})]$, i.e. $\overline{\alpha}$ is constant on any partition cell.
\end{theorem}

Importantly, the convex minimization program in \eqref{eq: ironed alpha} provides a convenient method to compute the ironed $\overline{\alpha}$. To deal with the majorization constraint in \eqref{eq: ironed alpha}, $g\succ\alpha$, we formulate $g$ as a function of $\alpha$ with a series of transfers from lower types to higher types, within the same agent.  Intuitively, this is reflects Theorem \ref{iffTtrans} and the fact that an orthogonal $T$ transform is equivalent to transfers between adjacent types of an agent. We know the transfer must be from a lower type to a  higher type by the inequalities that define $g\succ \alpha$. For $i\in N$, let $\lambda_i({\bf x})$ be the non-negative transfers for agent $i$ from her type $x_i$ to her next highest type with ${\bf x}_{-i}\in X_{-i}$. By convention, $\lambda_i(\overline{x}_i,{\bf x}_{-i})=0$ for all ${\bf x}_{-i}\in X_{-i}$. Define
\begin{equation}\label{g}
g(\mathbf{x})\,=\,\alpha(\mathbf{x})-\sum_{i\,\in\,N}\underline{\Delta}_i\lambda_i(\mathbf{x})/f(\mathbf{x}).
\end{equation}
We can confirm that $g$ does majorize $\alpha$: for any lower set $\mathbf{X}_-$,
\ba
\mathbb{E}\bigl[g(\mathbf{x})|\mathbf{x}\in\mathbf{X}_-\bigr]&=&\mathbb{E}\bigl[\alpha(\mathbf{x})-\sum_{i\in N}\underline\Delta_i \lambda_i(\mathbf{x})/f(\mathbf{x})\,|\,\mathbf{x}\in\mathbf{X}_-\bigr]\nonumber\\
&=&\mathbb{E}\bigl[\alpha(\mathbf{x})|\mathbf{x}\in\mathbf{X}_-\bigr]-\sum_{\mathbf{x}\,\in\,\overline{\partial}\mathbf{X}_-}\sum_{i\,\in\,N}\lambda_i(\mathbf{x})\nonumber\\
&\leq& \mathbb{E}\bigl[\alpha(\mathbf{x})|\mathbf{x}\in\mathbf{X}_-\bigr] \label{majMulti1}
\ea
where $\overline{\partial}\mathbf{X}_-$ denotes the set of extreme upper points of $\mathbf{X}_-$, i.e. $\mathbf{x}\in\overline{\partial}\mathbf{X}_-$ if there does not exist $\mathbf{y}\in\mathbf{X}_-$ with $\mathbf{y}\neq\mathbf{x}$ such that $\mathbf{x}\leq\mathbf{y}$.\footnote{ For instance, in the left panel of Figure \ref{uppersets}, the set of extreme upper points of $\mathbf{X}_-$ and $\mathbf{X}_-'$ coincide with their borders in the interior of $\mathbf{X}$ whereas the set of extreme points of $\mathbf{X}_-''$ is the collection of right-hand corners of each step in the border of $\mathbf{X}_-''$.}
Equation \eqref{majMulti1} holds with equality if all the $\lambda_i(\mathbf{x})$ vanish on $\overline{\partial}\mathbf{X}_-$. In particular, since $\lambda_i(\bar{x}_i,\mathbf{x}_{-i})=0$ for all $\mathbf{x}_{-i}\in\mathbf{X}_{-i}$ and $i\in N$, we have
\be\label{majMulti2}
\mathbb{E}\bigl[g(\mathbf{x})\bigr]=\mathbb{E}\bigl[\alpha(\mathbf{x})\bigr]
\ee
Together \eqref{majMulti1} and \eqref{majMulti2} show that $g$ defined in \eqref{g} majorizes $\alpha$.

Remarkably, the solution to \eqref{eq: ironed alpha} does not depend on the function $\phi$; for convenience, we will use $\phi(g) = g^2$. The $\lambda_i(\mathbf{x})$ can be obtained from the simple quadratic optimization program
\be\label{lambdastar}
\boldsymbol{\lambda}\,=\,
\argmin{\boldsymbol{\lambda}'\,:\,X\,\rightarrow\,\mathbb{R}^n_{\geq 0}}\mathbb{E}\bigl[(\alpha(\mathbf{x})-\sum_{i\,\in\,N}\underline{\Delta}_i\lambda'_i(\mathbf{x})/f(\mathbf{x}))^2\bigr]
\ee
which produces a unique $\boldsymbol{\lambda}$ and, hence, a unique $\overline{\alpha}$.
\addtocounter{example}{-1}
\begin{example}[continued]\label{partition}{\em
Recall that the minimal elements are given by
\begin{displaymath}
h\,=\,\biggl(\!\begin{array}{cc} 6-\lambda-\mu & \mu \\ \lambda & 6 \end{array}\!\biggr)
\end{displaymath}
with $\mu=\max(6-2\lambda,3-\hf\lambda)$ and $\lambda\in[0,6]$. Picking $\phi(x)=x^2$ in \eqref{eq: ironed alpha} yields $\lambda=\mu=2$ and
\begin{displaymath}
\overline{\alpha}\,=\,\biggl(\!\begin{array}{cc} 2 & 2 \\ 2 & 6 \end{array}\!\biggr)
\end{displaymath}
with ultramodular partition $\mathscr{P}=P_1\cup P_2$ where $P_1=\{(1,1),(1,2),(2,1)\}$ and $P_3=\{(2,2)\}$.}$\hfill\blacksquare$
\end{example}

\subsection{Restricted Least Squares}

In this section we outline an alternative method for computing the ironed $\alpha$ values. For $\alpha:\mathbf{X}\rightarrow\mathbb{R}$ not necessarily non-decreasing consider the restricted least-squares program
\begin{equation}\label{RLS}
\overline{\alpha}\,=\!\!\!
\min_{\substack{\rule{0pt}{10pt}g\,:\,\mathbf{X}\,\rightarrow\,\mathbb{R}_{\geq 0}\\\rule{0pt}{8pt}\text{s.t. $g(\mathbf{x})$ is non-decreasing}\\\rule{0pt}{8pt}\text{in each coordinate}}}
\hspace*{-5mm}||g(\mathbf{x})-\alpha(\mathbf{x})||^2
\end{equation}
where $|| h ||^2 = \mathbb{E} \big[h^2(\mathbf{x})\big]$ is the squared norm of $h$. The $\overline{\alpha}$ so obtained is identical to that obtained from \eqref{eq: ironed alpha}. To see this, write \eqref{RLS} as a saddle-point problem
\begin{displaymath}
\max_{{\bm \lambda}\,:\,\mathbf{X}\,\rightarrow\,\mathbb{R}_{\geq 0}}\,\,\min_{g\,:\,\mathbf{X}\,\rightarrow\,\mathbb{R}}\,\,\sum_{\mathbf{X}}(g(\mathbf{x})-\alpha(\mathbf{x}))^2dF(\mathbf{x})-
2\sum_{i\,\in\,\mathcal{N}}\,\sum_{\mathbf{X}}\lambda_i(\mathbf{x})\overline{\Delta}_ig_i(\mathbf{x})
\end{displaymath}
where the second term implements the non-decreasingness constraint. Changing the order of summation for this term and then taking first-order conditions yields
\begin{displaymath}
  \overline{\alpha}\,=\,\alpha(\mathbf{x})-\sum_{i\in\mathcal{N}}\underline{\Delta}_i\lambda_i(\mathbf{x})
\end{displaymath}
where ${\bm \lambda}^*(\mathbf{x})$ solves
\begin{equation}\label{one-way}
{\bm \lambda}^*\,=\,\argmax{{\bm \lambda}\,:\,\mathbf{X}\,\rightarrow\,\mathbb{R}_{\geq 0}}\,
\mathbb{E}[\alpha(\mathbf{x})^2-(\alpha(\mathbf{x})-\sum_{i\in\mathcal{N}}\underline{\Delta}_i\lambda_i(\mathbf{x}))^2]
\end{equation}
The solutions to \eqref{one-way} and \eqref{lambdastar} are identical.

For the one-dimensional case, \eqref{RLS} provides a computationally convenient alternative to Mussa and Rosen's (\citeyear{MussaRosen1978}) and Myerson's (\citeyear{Myerson1981}) methods.

\subsection{Continuous Types}\label{sec:contExt}

In this section, we show
that continuous types can be dealt with in much the same manner as the discrete-type case studied above.
Without loss of generality we assume $\mathbf{X}\,=\,[0,1]^N$. Let $(\mathbf{X},\,\mathscr{B},\,\{\mathscr{D}_n\}_{n\in\mathbb{N}},\,F)$ be a filtered probability space where $\mathscr{D}_n$ is the $\sigma$-algebra generated by dyadic cubes $d^N_n$ of sidelength $2^{-n}$, $\mathscr{B}\,=\,\sigma(\cup_n\mathscr{D}_n)$ is a Borel $\sigma$-algebra, and $F$ is an absolutely continuous distribution function.
\begin{definition}
Let $\mathscr{L}$ denote the collection of all lower subsets in $\mathscr{B}$ and let $g,h\,:\,\mathbf{X}\rightarrow\mathbb{R}$ be two integrable functions. We say $h$ majorizes $g$, denoted
$h\succ g$, iff
\begin{equation}
\int_{\mathbf{X}_-}(g(\mathbf{x})-h(\mathbf{x}))\mathrm{d}F(\mathbf{x})\,\geq\,0
\end{equation}
for all $\mathbf{X}_-\in\mathscr{L}$ with equality when $\mathbf{X}_-=\mathbf{X}$.
\end{definition}
Assume $\alpha$ is a bounded integrable function on $\mathbf{X}$. The conditional expectation of $\alpha$ on $\mathscr{D}_n$ is denoted $\alpha_n=\mathbb{E}[\alpha|\mathscr{D}_n]$. The $(\alpha_n)_{n\in\mathbb{N}}$ form a Doob martingale. Let $\overline{\alpha}_n$ be the ironed value that follows from Theorem \ref{form}.
\begin{theorem}\label{thm: continuous without access}
	$\overline{\alpha}_n$ has a subsequence converging almost everywhere to $\overline{\alpha}$ such that problems \eqref{problem: original problem} and \eqref{problem: relaxed problem} are equivalent.
\end{theorem}
The proof can be found in Appendix A where we also prove the stronger result that \textit{any} sequence $\overline{\alpha}_n$ converges to $\overline{\alpha}$ almost everywhere if the original $\alpha$ is continuous and has bounded variation.

For the continuous case, the geometric intuition behind multivariate majorization can be illustrated using the Gauss divergence theorem. Similar to the discrete case we define
\begin{displaymath}
  \overline{\alpha}(\mathbf{x})\,=\,\alpha(\mathbf{x})-\text{div}({\bm\lambda}({\bf x}))/f({\bf x})
\end{displaymath}
where ${\bm\lambda}({\bf x})\geq{\bm 0}$ for all $\mathbf{x}\in\mathbf{X}$ and $\text{div}({\bm\lambda}({\bf x}))=\sum_{i\in\mathcal{N}}\partial_{x_i}\lambda_i(\mathbf{x})$.
Consider any lower set $\mathbf{X}_-\subseteq\mathbf{X}$ with boundary $\partial\mathbf{X}_-$ that is the union of the upper boundary $\overline{\partial}\mathbf{X}_-$ and the lower boundary $\underline{\partial}\mathbf{X}_-$. The geometry of lower sets is such that the normal ${\bm n}({\bf x})$ to any point ${\bf x}$ on the upper boundary is positive, i.e. ${\bm n}({\bf x})\geq {\bm 0}$, whence ${\bm\lambda}({\bf x})\cdot{\bm n}({\bf x})\geq 0$. On the lower boundary we must have $x_i=\underline{x}_i$ for some $i\in N$, in which case the normal ${\bm n}({\bf x})$ is minus the $i$th unit vector and ${\bm\lambda}({\bf x})\cdot{\bm n}({\bf x})=0$ since $\lambda_i(\underline{x}_i,{\bf x}_{-i})=0$. By the Gauss divergence theorem we have
\ba
\mathbb{E}\bigl[\overline{\alpha}(\mathbf{x})\,|\,{\bf x} \in\mathbf{X}_-\bigr]&=&\mathbb{E}\bigl[\alpha(\mathbf{x})-\text{div}({\bm\lambda}(\mathbf{x}))/f(\mathbf{x})\,|\,{\bf x} \in X_-\bigr]\nonumber\\
&=&\mathbb{E}\bigl[\alpha(\mathbf{x})\,|\,{\bf x} \in\mathbf{X}_-\bigr]-\frac{1}{F(\mathbf{X}_-)}\int_{\overline{\partial}\mathbf{X}_-}{\bm\lambda}(\mathbf{x})\cdot{\bm n}(\mathbf{x})\nonumber\\
&\leq& \mathbb{E}\bigl[\alpha(\mathbf{x})\,|\,{\bf x} \in\mathbf{X}_-\bigr] \nonumber
\ea
with equality if ${\bm\lambda}({\bf x})\cdot{\bm n}({\bf x})=0$ on $\overline{\partial}X_-$. In particular, since $\lambda_i(\bar{x}_i,{\bf x}_{-i})=0$ for all ${\bf x}_{-i}\in X_{-i}$ and $i\in N$, we have $\mathbb{E}\bigl[\overline{\alpha}(\mathbf{x})\bigr]=\mathbb{E}\bigl[\alpha(\mathbf{x})\bigr]$.
To summarize, Definition \ref{multiVariate} readily extends to functions over continuous type spaces.

\begin{figure}[t]
\begin{center}
	\includegraphics[scale=.18]{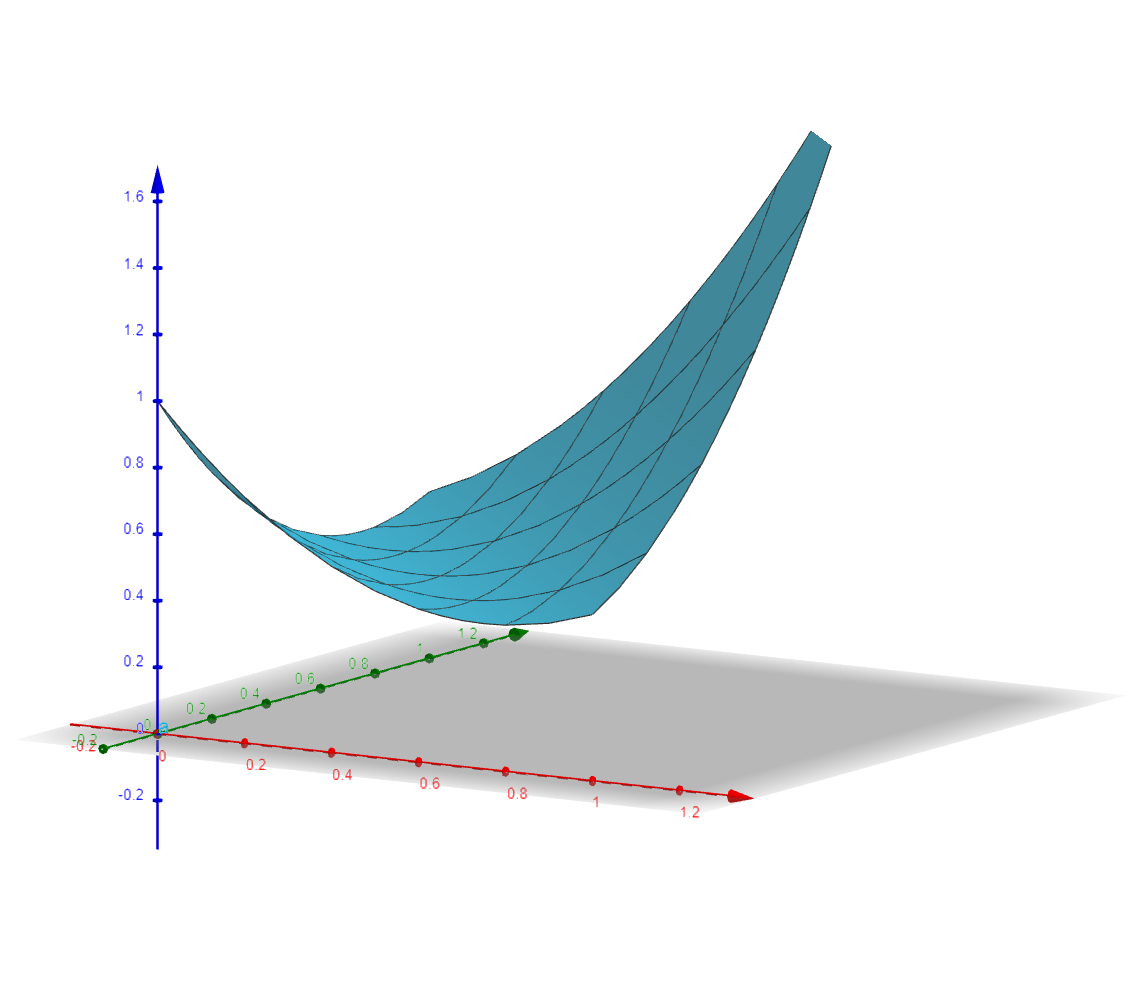}
	\hspace*{1mm}
	\includegraphics[scale=.18]{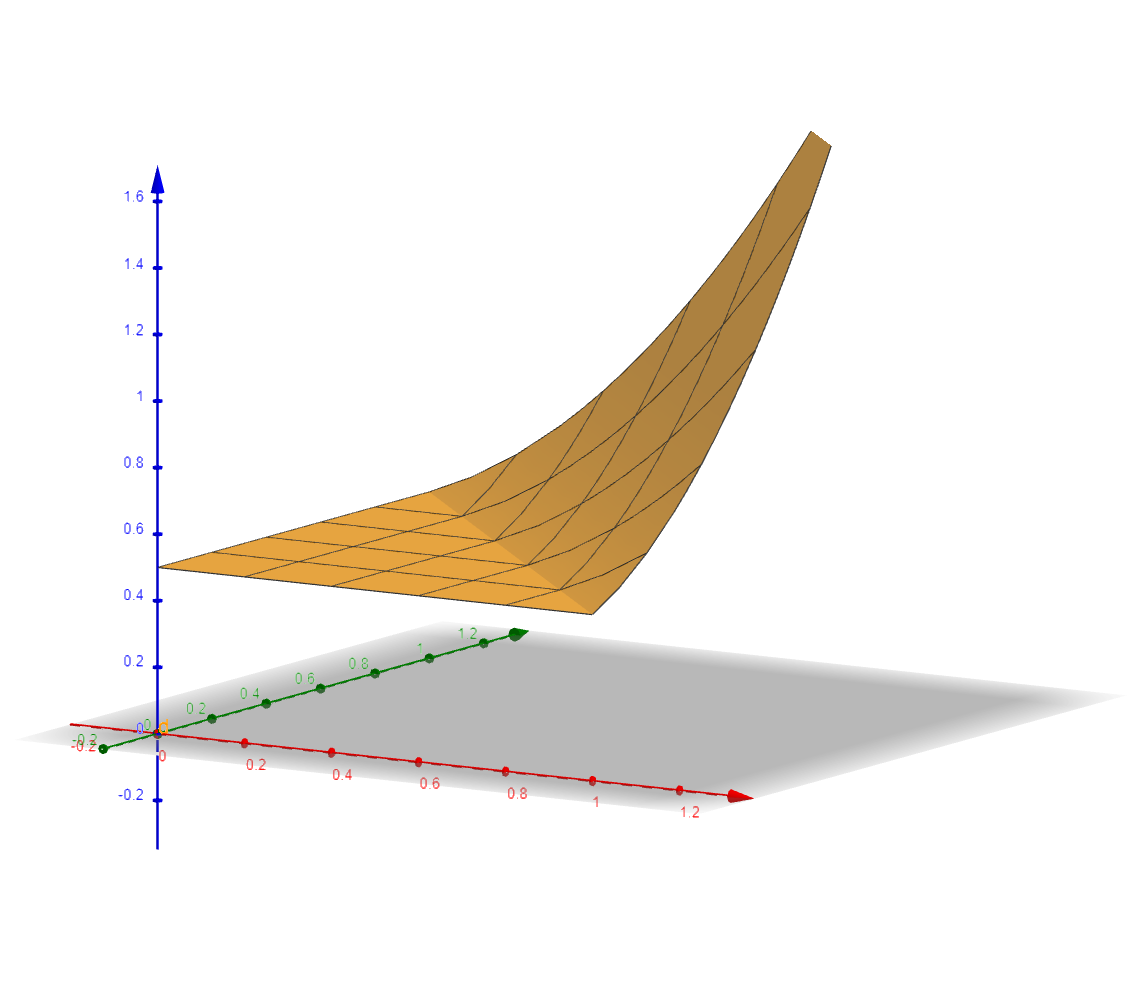}
\end{center}
\vspace*{-8mm}
	\caption{The $\alpha(\mathbf{x})$ of Example \ref{contex} (left panel) and its ironed version $\overline{\alpha}(\mathbf{x})$ (right panel).}
	\label{fig:example 2}
\vspace*{-3mm}
\end{figure}

\begin{example}\label{contex}{\em
Suppose $\alpha(x_1,x_2)=1-\deel{3}{2}(x_1+x_2)+(x_1+x_2)^2$ and types are uniformly distributed on $X=[0,1]^2$. The left panel of panel of Figure \ref{fig:example 2} shows $\alpha$. Its ironed version is given by
\begin{displaymath}
  \overline{\alpha}(x_1,x_2)\,=\,\Bigl\{\begin{array}{lll} \hf & \text{if} & x_1+x_2\,\leq\,1 \\[1mm] \alpha(x_1,x_2) & \text{if} & x_1+x_2\,\geq\,1\end{array}\Bigr.
\end{displaymath}
see the right panel of Figure \ref{fig:example 2}. It is readily verified that $\lambda_1(x,y)=\deel{1}{4}x_1(1-x_1-x_2)^2$ for $x_1+x_2\leq 1$ and zero otherwise and $\lambda_2(x_1,x_2)=\lambda_1(x_2,x_1)$ are such that
$\text{div}({\bm\lambda}(\mathbf{x}))=\alpha(\mathbf{x})-\overline{\alpha}(\mathbf{x})$.} $\hfill\blacksquare$
\end{example}

\section{Discriminatory Access Rights} \label{sec: discriminatory access rights}

In this section, we introduce access rights that allow the principal to ``screen out'' certain types. While our concept of majorization is most clearly demonstrated without screening, the ubiquity of such access rights in the real world makes it important that we show how our technique can accommodate them.\footnote{Access rights for public goods are also called ``exclusions'' in the literature. For example, see \cite{Cornelli1996,LedyardPalfrey1999,Hellwig2003} and \cite{Norman2004}. Our formulation below most closely follows \cite{Norman2004}.} For instance, mass-produced goods such as smart phones may be prohibitively costly for some buyers. Likewise, public goods are often restricted to a limited set of people. Besides all-or-nothing access rights used to screen out certain types we also consider probabilistic access rights that can be used to alleviate incentive constraints.

We show that the principal's problem
\be \label{constrainedMulti}
\max_{\substack{\rule{0pt}{10pt}(q,{\bm \eta})\,:\,\mathbf{X}\,\rightarrow\,\mathbb{R}_{\geq 0}\times [0,1]^{N}\\\rule{0pt}{8pt}\text{s.t. }\eta_i({\bf x})q({\bf x})\text{ is non-decreasing}\\\rule{0pt}{8pt}\text{in $x_i$ for all $i\in\mathcal{N}$}}}
\mathop{\mathbb{E}}\Bigl[q({\bf x})\sum_{i\,\in\,N}\eta_i({\bf x})\alpha_i({\bf x})-C(q({\bf x}))\Bigr]
\ee
can again be solved via multivariate majorization. To this end, we wish to manipulate each $\alpha_i$ to generate $\widetilde{\alpha}_i$ such that the following unconstrained problem is equivalent to \eqref{constrainedMulti}:
\be \label{problem: unconstrained problem with access right}
\max_{(\rule{0pt}{10pt}q,{\bm \eta})\,:\,\mathbf{X}\,\rightarrow\,\mathbb{R}_{\geq 0}\times [0,1]^{N}}
\mathop{\mathbb{E}} \Bigl[q(\mathbf{x})\sum_{i\,\in\,N}\eta_i(\mathbf{x})\widetilde{\alpha}_i(\mathbf{x})-C(q(\mathbf{x}))\Bigr]
\ee
We will need a notion of univariate majorization that applies to multidimensional functions.
\begin{definition}\label{uniVariate}
For $g:\mathbf{X}\rightarrow\mathbb{R}$, $h:\mathbf{X}\rightarrow\mathbb{R}$, $g$ \textbf{majorizes} $h$ \textbf{in coordinate} $i$, denoted $g\succ_i h$, if for all $\mathbf{s}\in\mathbf{X}$, $\mathbb{E}[g(\mathbf{x})|x_i\leq s_i,\mathbf{x}_{-i}=\mathbf{s}_{-i}]\leq\mathbb{E}[h(\mathbf{x})|x_i\leq s_i,\mathbf{x}_{-i}=\mathbf{s}_{-i}]$ with equality if $s_i=\overline{x}_i$.
\end{definition}
\noindent Now consider
\be\label{ironedMultiWith}
\widetilde{{\bm \alpha}}\,=\,\argmin{g_i\,:\,\mathbf{X}\,\rightarrow\,\mathbb{R}\atop g_i\,\succ_i\,\alpha_i}
\mathbb{E}\Bigl[\phi\big(\,\sum_{i\,\in\,N}\max(0,g_i(\mathbf{x}))\big)\Bigr]
\ee
for some strictly convex function $\phi:\mathbb{R}\rightarrow\mathbb{R}$. There is a trivial way in which the solution to \eqref{ironedMultiWith} is not unique, unlike the solution to \eqref{eq: ironed alpha}. If $\widetilde{\alpha}_i(\mathbf{x})<0$ for some $\mathbf{x}\in\mathbf{X}$, $i\in N$ then we could replace it with any other negative number. This multiplicity poses no problem since these cases are screened out and the principal's optimal decision is unique (see Lemma \ref{multiplicity} in Appendix A). Another difference with the previous section is that we cannot use level sets of $\widetilde{\alpha}({\bf x})=\sum_{i\in N}\widetilde{\alpha}_i({\bf x})$ to define the partition $\mathscr{P}$.
\begin{example}
{\em Suppose there are two agents with $X_1=X_2=\{x_1,x_2, x_3\}$ and types are uniformly distributed. Let
\begin{displaymath}
\alpha_1\,=\,\Biggl(\!\begin{array}{ccc} 0 & 9 & 0 \\ 1 & 10 & 1 \\ 2 & 11 & 2\end{array}\!\Biggr)\,\,\,\text{and}\,\,\,
\alpha_2\,=\,\Biggl(\!\begin{array}{ccc} 0 & 1 & 2 \\ 9 & 10 & 11 \\ 0 & 1 & 2\end{array}\!\Biggr)\,\,\,\text{so that}\,\,\,
\alpha\,=\,\alpha_1+\alpha_2\,=\,\Biggl(\!\begin{array}{ccc} 0 & 10 & 2 \\ 10 & 20 & 12 \\ 2 & 12 & 4 \end{array}\!\Biggr)
\end{displaymath}
For any strictly convex $\phi$, e.g. $\phi(x)=x^2$, Theorem \ref{form} yields
\begin{equation}\label{alpha}
    \overline{\alpha}\,=\,\Biggl(\!\begin{array}{ccc} 0 & 6 & 6 \\ 6 & 12 & 12 \\ 6 & 12 & 12 \end{array}\!\Biggr)
\end{equation}
which is the result without access rights, i.e. $\eta_i(\mathbf{x})=1$ for all $i\in\mathcal{N}$, $\mathbf{x}\in\mathbf{X}$. With access rights, the program in \eqref{ironedMultiWith} yields
\begin{equation}\label{alphatilde}
  \widetilde{\alpha}_1 \,=\,\Biggl(\!\begin{array}{ccc} 0 & 9 & 0 \\ 0 & 7 & 0 \\ 3 & 14 & 3 \end{array}\!\Biggr)\,\,\,\text{and}\,\,\,
  \widetilde{\alpha}_2 \,=\,\Biggl(\!\begin{array}{ccc} 0 & 0 & 3 \\ 9 & 7 & 14 \\ 0 & 0 & 3 \end{array}\!\Biggr)\,\,\,\text{so that}\,\,\,
  \widetilde{\alpha}\,=\,\Biggl(\!\begin{array}{ccc} 0 & 9 & 3 \\ 9 & 14 & 14 \\ 3 & 14 & 6 \end{array}\!\Biggr)
\end{equation}
Based on level sets of $\widetilde{\alpha}$, the partition would consist of five cells: two singletons, two sets of size two, and a set of size three. However, this partition violates the majorization requirement that $\widetilde{\alpha}$ has the same expected value as $\alpha$ on each cell.

The correct ultramodular partition can be obtained from the program in \eqref{ironedMultiWith}, which can be executed by parameterizing $g_i(\mathbf{x})=\alpha_i(\mathbf{x})-\underline{\Delta}_i\lambda_i(\mathbf{x})$, with $\lambda_i(\mathbf{x})\geq 0$ and $\lambda_i(\overline{x}_i,\mathbf{x}_{-i})=0$ for $i=1,2$ and $\mathbf{x}\in\mathbf{X}$, and then minimizing over the $\lambda_i$s. This yields
\begin{displaymath}
\lambda_1\,=\,\Biggl(
\begin{tikzpicture}[baseline=($(0.base)!.6!(2.base)$)]
  \matrix[matrix of math nodes,row sep = 1pt,column sep = 5pt] (m)
  {
    0  & 0 & 0\\
    1  & 3 & 1\\
    0  & 0 & 0\\
  };
  \begin{pgfonlayer}{background}
    \node[inner sep=-1pt,fit=(m-1-1)]          (0)   {};
    \node[inner sep=-1pt,fit=(m-2-1)]          (1)   {};
    \node[inner sep=-1pt,fit=(m-3-1)]          (2)   {};
    \node[inner sep=-1pt,fit=(m-1-2)]          (3)   {};
    \node[inner sep=-1pt,fit=(m-2-2)]          (4)   {};
    \node[inner sep=-1pt,fit=(m-2-3)]          (5)   {};
    \node[inner sep=-1pt,fit=(m-3-3)]          (6)   {};
    \node[inner sep=-1pt,fit=(m-3-2)]          (7)   {};
    \draw[rounded corners,draw=green!50!white,fill=green!50!white,inner sep=3pt,fill opacity=0.1] (1.north west) |- (2.south east) |- (2.east) |- (1.north) -- cycle;
    \draw[rounded corners,draw=red!50!white,fill=red!50!white,inner sep=3pt,fill opacity=0.1] (4.north west) |- (7.south east) |- (7.east) |- (4.north) -- cycle;
    \draw[rounded corners,draw=red!50!white,fill=red!50!white,inner sep=3pt,fill opacity=0.1] (5.north west) |- (6.south east) |- (6.east) |- (5.north) -- cycle;
  \end{pgfonlayer}
\end{tikzpicture}
\Biggr)
\hspace*{2cm}
\lambda_2\,=\,\Biggl(
\begin{tikzpicture}[baseline=($(0.base)!.6!(5.base)$)]
  \matrix[matrix of math nodes,row sep = 1pt,column sep = 5pt] (n)
  {
    0  & 1 & 0\\
    0  & 3 & 0\\
    0  & 1 & 0\\
  };
  \begin{pgfonlayer}{background}
    \node[inner sep=-1pt,fit=(n-1-1)]          (0)   {};
    \node[inner sep=-1pt,fit=(n-1-2)]          (1)   {};
    \node[inner sep=-1pt,fit=(n-1-3)]          (2)   {};
    \node[inner sep=-1pt,fit=(n-2-1)]          (3)   {};
    \node[inner sep=-1pt,fit=(n-2-2)]          (4)   {};
    \node[inner sep=-1pt,fit=(n-3-2)]          (5)   {};
    \node[inner sep=-1pt,fit=(n-3-3)]          (6)   {};
    \node[inner sep=-1pt,fit=(n-2-3)]          (7)   {};
    \draw[rounded corners,draw=green!50!white,fill=green!50!white,inner sep=3pt,fill opacity=0.1] (1.north west) |- (2.south east) |- (2.east) |- (1.north) -- cycle;
    \draw[rounded corners,draw=red!50!white,fill=red!50!white,inner sep=3pt,fill opacity=0.1] (4.north west) |- (7.south east) |- (7.east) |- (4.north) -- cycle;
    \draw[rounded corners,draw=red!50!white,fill=red!50!white,inner sep=3pt,fill opacity=0.1] (5.north west) |- (6.south east) |- (6.east) |- (5.north) -- cycle;
  \end{pgfonlayer}
\end{tikzpicture}
\Biggr)
\end{displaymath}
For agent $i$ we fix ${\bf x}_{-i}$ and find the largest $x_i$ (if any) for which $\lambda_i(x_i,\mathbf{x}_{-i})>0$ as well as the largest $x'_i<x_i$ for which $\lambda_i(x_i,\mathbf{x}_{-i})=0$ to form the sets, $\{(x,\mathbf{x}_{-i})|x'_i\leq x\leq x_i\}$, which are highlighted above.  We do this for all agents and then \q{overlay} the individual sets to find the partition.
Profiles that do not belong to any of the highlighted sets form singletons on which $\widetilde{\alpha}({\bf x})=\alpha({\bf x})$. For the above example, the correct partition is: $\mathscr{P}=P_1\cup P_2\cup P_3\cup P_4$ where $P_1=\{(0,0)\}$,  $P_2=\{(0,1),(0,2)\}$, $P_3=\{(1,0),(2,0)\}$, and $P_4=\{(1,1),(1,2),(2,1),(2,2)\}$. Now $\widetilde{\alpha}$ has the same expected values as $\alpha$ on each of the cells.}$\hfill\blacksquare$
\end{example}
\begin{theorem} \label{thm: iron with access rights}
	The solution to the full problem \eqref{constrainedMulti} is
	\be
	q^*({\bf x})\,=\,C'^{-1}\bigl(\,\sum_{i\,\in\,N}\max(0,\widetilde{\alpha}_i({\bf x}))\bigr)
	\ee
	and with $\widetilde{{\bm \alpha}}$ is associated a ultramodular partition $\mathscr{P}$ of $X$. For ${\bf x}\in X$, let $P({\bf x})$ be the cell of $\mathscr{P}$ containing ${\bf x}$ and let $P_i({\bf x})=\{{\bf y}\in P({\bf x})|{\bf y}_{-i} = {\bf x}_{-i}\}$. The optimal access rights ${\bm \eta}^*$ is given by
	\begin{displaymath}
	\eta^*_i({\bf x})=
	\left\{
	\begin{array}{lll}
	0 & \text{\em if} & \widetilde{\alpha}_i({\bf x})<0\\[1mm]
	\eta_i^0({\bf x}) & \text{\em if} & \widetilde{\alpha}_i({\bf x})=0\\[1mm]
	1 & \text{\em if} & \widetilde{\alpha}_i({\bf x})>0
	\end{array}
	\right.
	\end{displaymath}
	where $\eta_i^0({\bf x})$ is such that $q^*({\bf x})\eta_i^0({\bf x})$ is constant on $P_i({\bf x})$ and non-decreasing in $x_i$.\footnote{In detail, (i) if there exists ${\bf y}\in P_i({\bf x})$ such that $\widetilde{\alpha}_i({\bf y})\neq 0$ then $q^*({\bf x}) \eta_i^0({\bf x}) = q^*({\bf y}) \eta_i^*({\bf y})$, (ii) otherwise $q^*({\bf x})\eta_i^0({\bf x})=q^*({\bf y})\eta_i^0({\bf y})$ for all ${\bf y}\in P_i({\bf x})$ and $q^*(\underline {\bf y}) \eta_i^*(\underline {\bf y}) \leq q^*({\bf x}) \eta_i^0({\bf x}) \leq q^*(\overline{\bf y}) \eta_i^*(\overline{\bf y})$ for any $\underline{ y}_i \leq x_i\leq \overline{y}_i$ with $\widetilde{\alpha}_i(\underline {\bf y})\neq 0$ and $\widetilde{\alpha}_i(\overline {\bf y})\neq 0$.}
\end{theorem}
\addtocounter{example}{-1}
\begin{example}[continued]{\em
Unlike $\overline{\alpha}$, the solution $\widetilde{\alpha}$ that follows from \eqref{ironedMultiWith} is not necessarily constant on a cell, cf. \eqref{alpha} and \eqref{alphatilde} above. Incentive compatibility is maintained by choosing appropriate access rights, i.e.
\begin{displaymath}
  \eta_1\,=\,\Biggl(\begin{array}{ccc} 0 & 1 & 0 \\[1mm] \deel{1}{3} & 1 & \deel{3}{7} \\[1mm] 1 & 1 & 1 \end{array}\Biggr)
\end{displaymath}
and $\eta_2$ is the transpose of $\eta_1$. This example shows that besides all-or-nothing access rights that can be used to screen out agents, probabilistic access rights can be beneficial in that they allow the ironed $\widetilde{\alpha}_i$ to more closely track the original $\alpha_i$ when the latter are non-monotonic.}$\hfill\blacksquare$
\end{example}

\section{Applications}

In this section we detail three applications of our concept of multivariate majorization. First we discuss two examples of principal-agents problems with interdependent agent types. Next we show how applying the theory of multivariate majorization to cumulative distribution functions provides a natural extension of second-order stochastic dominance, providing a useful tool for analyzing multivariate risk.

\subsection{Optimal Mechanisms For Mass-Produced Goods}

Any seller of a mass-market good faces the challenge of designing a single good to appeal to a large number of diverse consumers. The decision of interest is not how many copies of a good to make but how to best design the template used to generate copies. In particular, the seller chooses a single quality level for the good to be enjoyed commonly by all consumers of the good. In addition, the seller may restrict access to the good and collect transfers.

Virtually any product sold as identical goods to more than one person will fit into this framework. For example, producers of popular movies, books, television programming and music rely on wide appeal to generate profits, rather than finding the ideal fit for each customer. Schools must contend with how to design lectures and deliver courses and, as technology relaxes physical constraints to the learning environment, whom to exclude, if anyone. Mass-produced goods such as furniture, electronics, and some food and drinks have the same characteristic. Most restaurants offer consistent menus over time as customers flow through, certainly in the case of national and global chains. Many non-profit outfits such as library, schools, museums, theatres and orchestras, and public parks similarly satisfy large groups with limited offerings.

We assume that a buyer's valuation for the good depends on her own signal as well as the signals of all other buyers, i.e. valuations are interdependent. In particular, buyer $i$'s valuation for the good given type profile ${\bf x}\in X$ is $v_i({\bf x})q({\bf x})$, where $v_i:\mathbf{X}\rightarrow\mathbb{R}$ is non-decreasing in $x_i$ for all ${\bf x}_{-i}\in\mathbf{X}_{-i}$. Buyer $i$'s payoff, given choices $(q,{\bm \eta},{\bm t})$ by the seller and type profile ${\bf x}\in\mathbf{X}$ is $u_i(q,{\bm \eta},{\bm t}; {\bf x})=v_i({\bf x})q({\bf x})\eta_i({\bf x})-t_i({\bf x})$.

\begin{figure}[t]
\begin{center}
	\includegraphics[width=8cm,height=7cm]{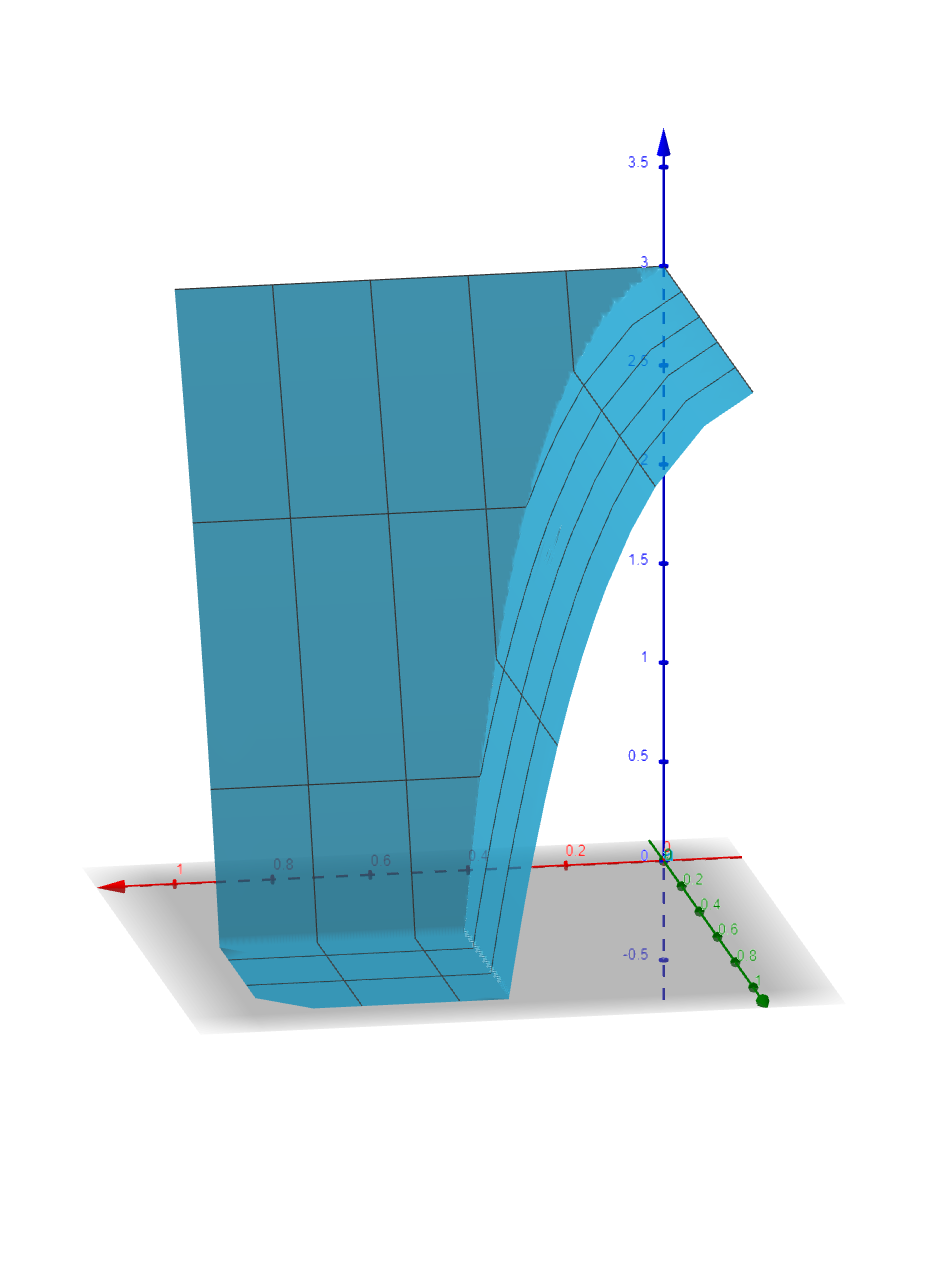}
	\hspace*{-13mm}
	\includegraphics[width=8cm,height=7cm]{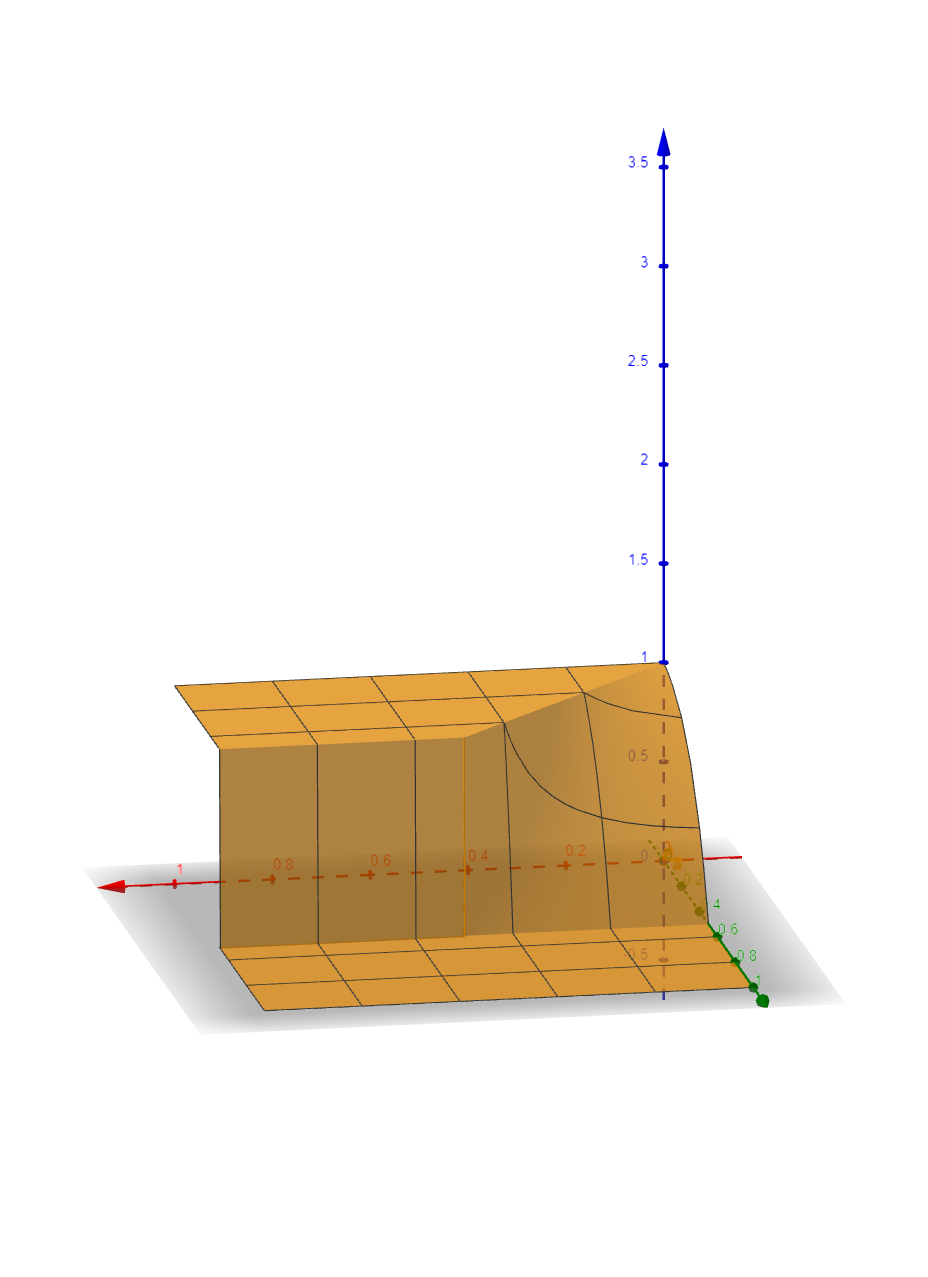}
\end{center}
\vspace*{-15mm}
	\caption{The optimal quality (left panel) and access rights (right panel) of Example \ref{ex:cont}.}
	\label{fig:exCont}
\end{figure}

The seller's problem is to choose a mechanism to maximize the expected sum of transfers from buyers net of the expected cost. The cost of providing quantity $q$ is $C(q)$, which is assumed to be an increasing convex (and differentiable) function with $C(0)=C'(0)=0$. Due to the revelation principle, we can focus on (incentive compatible) direct mechanisms: $(q,{\bm \eta},{\bm t})$ where (i) $q:\mathbf{X}\rightarrow \mathbb{R}_{\geq 0}$ maps type profiles into quantity choices; (ii) ${\bm \eta}:\mathbf{X}\rightarrow [0,1]^N$ maps type profiles into access rights; and (iii) ${\bf t}:\mathbf{X}\rightarrow \mathbb{R}^N$ maps type profiles into transfers. The seller's problem is therefore to choose a direct mechanism $(q,{\bm \eta},{\bm t})$ to maximize
\bd
\mathbb{E}\,\Bigl[\,\sum_{i\,\in\,N} t_i({\bf x})-C(q({\bf x}))\,\Bigr]
\ed
subject to (ex post) incentive compatibility and individual rationality. Using standard arguments, we can solve for the transfers
\begin{displaymath}\label{transfers}
t_i({\bf x})\,=\,v_i({\bf x})q({\bf x})\eta_i({\bf x})-\int_0^{x_i}v'_i(s_i,{\bf x}_{-i})q(s_i,{\bf x}_{-i})\eta_i(s_i,{\bf x}_{-i})ds_i
\end{displaymath}
where $v'_i(\mathbf{x})=\partial_{x_i}v_i(x_i,\mathbf{x}_{-i})$ to rewrite the seller's profit as
\begin{displaymath}\label{profit}
\Pi(\bm{M\!R},q,{\bm \eta})\,=\,\mathbb{E}\,\Bigl[\,q({\bf x})\sum_{i\,\in\,N}M\!R_i({\bf x})\eta_i({\bf x})-C(q({\bf x}))\,\Bigr]
\end{displaymath}
where $\bm{M\!R}=\{M\!R_1,\ldots,M\!R_N\}$ and, for $i\in\mathcal{N}$,
\begin{displaymath}\label{mr}
M\!R_i({\bf x})\,=\,v_i({\bf x})-\frac{1-F_i(x_i)}{f_i(x_i)}\,v'_i({\bf x})
\end{displaymath}
The seller's problem is thus to choose the quality, $q({\bf x})$, and access rights, ${\bm \eta}({\bf x})$, to maximize $\Pi(\bm{M\!R},q,{\bm \eta})$ such that $\eta_i({\bf x})q({\bf x})$ is non-decreasing in $x_i$ for all $i\in N$, ${\bf x}\in X$ and set transfers, ${\bm t}({\bf x})$, as above. Let $\Pi^*$ denote the seller's profits when using the optimal mechanism $(q^*,{\bm \eta}^*,{\bf t}^*)$.
\begin{example}[Probabilistic access rights with continuous types]\label{ex:cont}
{\em
For $i=1,2$, let buyers' value functions be given by $v_i(x_i,x_{-i})=3-4(x_i+x_{-i})-4(x_i+x_{-i})^2$ where the types are uniformly distributed on $\mathbf{X}=[0,1]^2$. The sum of marginal revenues
\begin{displaymath}
  M\!R(x_1,x_2)\,=\,14+4(x_1+x_2)-16(x_1+x_2)^2
\end{displaymath}
is decreasing for higher types and $\mathbb{E}[M\!R(x_1,x_2)]=-\deel{2}{3}$. As a result, Theorem \ref{form} yields $q^*(x_1,x_2)=0$, i.e. without access rights the seller's optimal choice is to bunch all buyers and offer zero quality independent of their types.

With access rights, however, the optimal quality that follows from Theorem \ref{thm: iron with access rights} is
\begin{displaymath}
  q^*(x_1,x_2)\,=\,\max\bigl(0,\frac{(3+2x_1)(1-2x_1)}{1-x_1},\frac{(3-2x_2)(1-2x_2)}{1-x_2}\,\bigr)
\end{displaymath}
which bunches buyer pairs according to $\min(x_1,x_2)$, i.e. the minimum of their types. When both buyers have above-average types, $\min(x_1,x_2)\geq\hf$, the seller offers them zero quality. Otherwise, they get offered a positive quality that is higher the lower is $\min(x_1,x_2)$. While the optimal quality $q(x_1,x_2)$ is constant in a buyer's type when the buyer has the higher type it is strictly \textit{decreasing} in a buyer's type when the buyer has the lower type. To maintain incentive compatibility the seller optimally uses probabilistic access rights
\begin{displaymath}
  \eta^*_1(x_1,x_2)\,=\,\left\{\begin{array}{lll}
  1 & \text{if} & x_1\,\geq\,x_2\,\,\text{and}\,\,x_2\,\leq\,\hf\\[1mm]
  \frac{(3+2x_2)(1-2x_2)(1-x_1)}{(3+2x_1)(1-2x_1)(1-x_2)} & \text{if} & x_1\,\leq\,x_2\,\,\text{and}\,\,x_2\,\leq\,\hf\\[1mm]
  0 & \text{if} & x_2\,>\,\hf \end{array}\right.
\end{displaymath}
and $\eta^*_2(x_1,x_2)=\eta^*_1(x_2,x_1)$. The optimal quality and access rights are shown in the left and right panels of Figure \ref{fig:exCont}.}
\end{example}

Besides the interpretation of the seller choosing design quality, one can also view this model as the private provision of a public good with private information, with the seller choosing the quantity of the public good to provide. The analysis of this problem in a mechanism design framework goes back at least as far as \cite{GrovesLedyard1977}. Classically, the decision is binary, i.e. to provide the good or not, and participation is compulsory, i.e. the good is non-excludable. The relevant question is whether the good is provided efficiently \citep[see][]{GuthHellwig1986}. More recently, exclusions and continuous levels of the good have been allowed \citep[see][]{Cornelli1996,LedyardPalfrey1999,Hellwig2003,Norman2004}. Preferences in these papers are purely private valued.

More broadly, there is a separate literature exploring the implementation of social choice functions when agents have interdependent preferences \citep[see][]{BergemannMorris2009, OllarPenta2017, OllarPenta2021}. These papers are not concerned with optimal mechanisms per se but rather with exploring how a principal can implement particular equilibrium outcomes given various assumptions of the principal's knowledge of agents' first- and higher-order beliefs.

\subsection{Multi-Agent Contracting}

Consider a contracting relationship where the principal chooses a duration, $d$, during which agents must exert effort. Examples include a factory deciding the number of hours in a workday or a coach choosing the length of a football practice. For $i\in\mathcal{N}$, agent $i$'s cost of exerting effort for a total amount of time $d$ equals $c_i(\mathbf{x})d$ where the marginal cost $c_i(\mathbf{x})$ is decreasing in the agent's own type $x_i$ and may depend on others' types. For example, training an hour with professional footballers may be unusually exhausting. The principal's payoff is $y(d)-\sum_{i\in\mathcal{N}}t_i$ where the production function $y$ has decreasing marginal returns and $t_i$ is the transfer to agent $i$. For $i\in\mathcal{N}$, agent $i$'s payoff is $t_i-c_i(\mathbf{x})d$.

As in the previous section, we can solve out the transfers using standard arguments and write the principal's problem as
\begin{equation} \label{contract}
\max_{\substack{\rule{0pt}{10pt}d\,:\,\mathbf{X}\,\rightarrow\,\mathbb{R}_{\geq 0}\\\rule{0pt}{8pt}\text{s.t. $d(\mathbf{x})$ is non-decreasing}\\\rule{0pt}{8pt}\text{in each coordinate}}}
\hspace*{-5mm}\mathbb{E}\bigl[y(d(\mathbf{x}))-d(\mathbf{x})MC(\mathbf{x})\bigr]
\end{equation}
with
\begin{displaymath}
  MC(\mathbf{x})\,=\,\sum_{i\,\in\,\mathcal{N}}c_i(\mathbf{x})-\frac{1-F_i(x_i)}{f_i(x_i)}c'_i(\mathbf{x})
\end{displaymath}
and $c'_i(\mathbf{x})=\partial_{x_i}c_i(\mathbf{x})$. With continuous types the principal's optimal choice is
\begin{displaymath}
  d^*(\mathbf{x})\,=\,{y'}^{-1}(-(\overline{-MC}(\mathbf{x})))
\end{displaymath}
where $\overline{-MC}(\mathbf{x})$ follows from \eqref{eq: ironed alpha}. Discrete types can readily be handled as well by replacing $MC(\mathbf{x})$ with its discrete version
\begin{displaymath}
  MC(\mathbf{x})\,=\,\sum_{i\,\in\,\mathcal{N}}c_i(\mathbf{x})-\frac{1-F_i(x_i)}{f_i(x_i)}\overline{\Delta}c_i(\mathbf{x})
\end{displaymath}
\vspace*{-8mm}

\begin{example} {\em
Suppose $N=2$ and, for $i=1,2$, the cost functions are $c_i(x_i,x_{-i})=(3-x_i)(1+x_{-i})$ where the types are uniformly distributed and $X_i=\{0,1,2\}$. The sum of agents' marginal costs
\begin{displaymath}
  -MC\,=\,-\Biggl(\!\begin{array}{ccc} 10 & 13 & 16 \\ 13 & 12 & 11 \\ 16 & 11 & 6\end{array}\!\Biggr)
\end{displaymath}
is not everywhere non-decreasing. The ironed version
\begin{displaymath}
  \overline{-MC}\,=\,-\Biggl(\!\begin{array}{ccc} 13\deel{3}{5} & 13\deel{3}{5} & 13\deel{3}{5} \\[1mm] 13\deel{3}{5} & 12 & 11 \\[1mm] 13\deel{3}{5} & 11 & 6\end{array}\!\Biggr)
\end{displaymath}
yields the optimal duration ${y'}^{-1}(-(\overline{-MC}(\mathbf{x})))$.}
\end{example}

\subsection{Decisionmaking Under Multivariate Risk}

Second-order stochastic dominance is a powerful tool to rank uncertain prospects for a broad class of preferences. The concept was developed over half a century ago (e.g. Hadar and Russell, \citeyear{HadarRussell1969}; Hanoch and Levy, \citeyear{HanochLevy1969}; Rothschild and Stiglitz, \citeyear{RothschildStiglitz1970}) and has since been applied in economics, finance, decision theory, mathematical programming, stochastic processes, etc. Recall that distribution function $G:[0,1]\rightarrow[0,1]$ second-order stochastically dominates the distribution function $F:[0,1]\rightarrow[0,1]$ if
\begin{displaymath}
  \int_0^xG(y)dy\,\leq\,\int_0^xF(y)dy
\end{displaymath}
for all $x\in[0,1]$ with equality when $x=1$. (The latter ensures that $\mathbb{E}[X]=\mathbb{E}[Y]$ as can be verified by partial integration.) This is the \textit{exact} definition of univariate majorization. In other words, second-order stochastic dominance is simply univariate majorization applied to a restricted class of (distribution) functions.

Second-order stochastic dominance is well-known to be related to mean-preserving spreads. Random variable $X$ is a mean-preserving spread of random variable $Y$ if $X=Y+\varepsilon$ with $\mathbb{E}[\varepsilon|Y]=0$. For the densities $f$ and $g$ associated with $X$ and $Y$ respectively, a mean-preserving spread implies
\begin{displaymath}
  f(x)\,=\,\int_0^1t(x,y)g(y)dy
\end{displaymath}
where $t:[0,1]\times[0,1]\rightarrow[0,1]$ is a \textit{sweeping operator}, e.g. \cite{RochetChone1998}, that satisfies
\begin{displaymath}
  \int_0^1t(x,y)dx\,=\,1\,\,\,\,\,\text{and}\,\,\,\,\,\int_0^1xt(x,y)dx\,=\,y
\end{displaymath}
for $y\in[0,1]$, with boundary conditions $t(x,0)=\delta(x)$ and $t(x,1)=\delta(1-x)$ since the endpoints cannot be swept in a mean-preserving manner (here $\delta(\cdot)$ denotes the Dirac delta function).
\begin{theorem}\label{sweep2DS}
Let $F$ and $G$ be the distributions associated with $X$ and $Y$ respectively. Then $X$ is a mean-preserving spread of $Y$ if and only if
\begin{displaymath}
  F(x)\,=\,\int_0^1T(x,y)G(y)dy
\end{displaymath}
with $T(x,y)$ a doubly-stochastic operator, i.e. $\int_0^1T(x,y)dx=\int_0^1T(x,y)dy=1$ for $x,y\in[0,1]$.
\end{theorem}
We thus can state the following variant of a classic result due to Rothschild and Stiglitz.
\begin{theorem}[Rothschild and Stiglitz, \citeyear{RothschildStiglitz1970}]\label{RS}
The following statements are equivalent:
\begin{itemize}\addtolength{\itemsep}{-1mm}
\vspace*{-2mm}
\item[1.] $G(x)$ second-order stochastically dominates $F(x)$.
\item[2.] $F(x)=\int_0^1 T(x,y)G(y)dy$ where $T(x,y)$ is a doubly-stochastic operator.
\item[3.] $\int_{[0,1]}u(x)g(x)dx\geq\int_{[0,1]}u(x)f(x)dx$ for any concave utility function $u(x)$.
\end{itemize}
\end{theorem}
Most of the literature has focused on univariate risk where wealth is the only random variable.\footnote{See e.g. the discussion in Chapter 17 of \cite{Levy2016} who notes that papers that consider multivariate risk focus only on first-order stochastic dominance.} However, many decisions involve multiple sources of risk, e.g. when individuals care about their health and their wealth.

First-order stochastic dominance can readily be generalized to multidimensional settings using the lower sets introduced above. Let $f(\mathbf{x})$ and $g(\mathbf{x})$ denote the densities associated with
$F(\mathbf{x})$ and $G(\mathbf{x})$ respectively.  Then $G$ first-order dominates $F$ if
\begin{displaymath}
  \int_{\mathbf{X}_-}g({\bf x})d\mathbf{x}\,\leq\,\int_{\mathbf{X}_-}f({\bf x})d\mathbf{x}
\end{displaymath}
for any lower set $\mathbf{X}_-\subseteq\mathbf{X}$ with equality when $\mathbf{X}_-=\mathbf{X}$. However, there is no definition in the literature for multivariate second-order stochastic dominance.
We next provide such a definition.

To a multivariate distribution function $F:[0,1]^N\rightarrow[0,1]$ we associate the complement of the survival function
\begin{displaymath}
  \overline{F}(\mathbf{x})\,=\,1-\int_{\mathbf{z}\,\geq\,\mathbf{x}}dF(\mathbf{z})
\end{displaymath}
which is non-decreasing in each argument with $\overline{F}({\bm 0})=0$ and $\overline{F}(\mathbf{x})=1$ if one or more of the $x_i=1$. Note that for the univariate case we have $\overline{F}(x)=F(x)$. Our definition for multivariate second-order stochastic dominance is simply that of multivariate majorization.
\begin{definition}\label{def:SOSD}
$G(\mathbf{x})$ second-order stochastically dominates $F(\mathbf{x})$ if
\begin{equation}\label{Xplus2}
  \int_{\mathbf{X}_-}\overline{G}({\bf x})d\mathbf{x}\,\leq\,\int_{\mathbf{X}_-}\overline{F}({\bf x})d\mathbf{x}
\end{equation}
for any lower set $\mathbf{X}_-\subseteq[0,1]^N$ with equality when $\mathbf{X}_-=[0,1]^N$.
\end{definition}
Let $2^{\mathcal{N}}$ denote the set of all non-empty subsets of $\mathcal{N}$.  An element of $2^{\mathcal{N}}$ is a set of labels $S=\{i_1,\ldots,i_{\ell}\}$ where $\ell=1,\ldots,N$ and the $i_k$ for  $k=1,\ldots,\ell$ are different elements of $\mathcal{N}$. Let $\overline{G}_S({\bf x}_S)$ equal $\overline{G}({\bf x})$ with the $i$-th argument set to zero if $i\not\in S$. Likewise, let $u_S({\bf x}_S)$ equal $u({\bf x})$ with the $i$-th argument set to zero if $i\not\in S$. Finally, let $D_S=\prod_{i\in S}\partial_{x_i}$.
\begin{definition}
A utility function is \textit{ortho-concave} if $D_Su({\bf x}_S)$ is non-increasing in each argument for all $S\in2^{\mathcal{N}}$.
\end{definition}
Examples of ortho-concave functions include Cobb-Douglas utility functions $u(\mathbf{x})=x_1^{\beta_1}\cdots x_N^{\beta_N}$ with $\beta_i\in[0,1]$ for $i\in\mathcal{N}$.\footnote{Note that $D_S u({\bf x})\geq 0$ for $S=\mathcal{N}$ and $D_S u({\bf x})=0$ for $S\subset\mathcal{N}$ since the Cobb-Douglas utility vanishes if one or more arguments are zero.} Note that $u(\mathbf{x})$ is concave only when in addition $\sum_{i\in\mathcal{N}}\beta_i\leq 1$.
\begin{theorem}\label{theorem:SOSD}
The following statements are equivalent:
\begin{itemize}\addtolength{\itemsep}{-1mm}
\vspace*{-2mm}
\item[1.] $G(\mathbf{x})$ second-order stochastically dominates $F(\mathbf{x})$.
\item[2.] $\overline{F}(\mathbf{x})=\int_{[0,1]^N}T(\mathbf{x},\mathbf{y})\overline{G}(\mathbf{y})d\mathbf{y}$ where $T(\mathbf{x},\mathbf{y})$ is an orthogonal doubly-stochastic operator.
\item[3.] $\int_{[0,1]^N}u(\mathbf{x})g(\mathbf{x})d\mathbf{x}\geq\int_{[0,1]^N}u(\mathbf{x})f(\mathbf{x})d\mathbf{x}$ for any ortho-concave utility function $u(\mathbf{x})$.
\end{itemize}
\end{theorem}

\section{Conclusion}

We develop a constructive approach to multidimensional ironing via a novel concept of multivariate majorization. Using this new technique, we characterize the optimal mechanism for a class of principal-agents problems when agents' values are interdependent. We show how the interdependency of agents' values creates multidimensional incentive-compatibility constraints on the principal's problem. These constraints cannot be addressed using the univariate ``ironing'' method developed by \cite{MussaRosen1978} and \cite{Myerson1981}. Instead,
we use multivariate majorization to jointly iron buyers' preferences. The resulting preferences are sufficiently well behaved that the incentive constraints no longer strictly bind allowing the principal to effectively ignore them.

While we mostly focus on the case where all agents are guaranteed access to the good regardless of their type, we show that our approach extends directly to the case of excludable access. In so doing, we demonstrate that exclusions and random access rights are important ways for the monopolist to raise profits. The former allows the monopolist to exclude buyers with negative marginal revenue and the latter allows the monopolist to fine tune incentives, both without manipulating the overall choice of quality for the market.

We demonstrate that the same multivariate majorization technique applies to discrete and continuous type spaces, unlike the univariate ironing methods developed by \cite{MussaRosen1978} and \cite{Myerson1981}. We further show that the ironed function can be obtained via a restricted least-squares problem, both in the univariate and multivariate case.

Restricting the class of multivariate functions to cumulative distribution functions, our multivariate majorization technique provides a multidimensional extension of second-order stochastic dominance. This allows us to generalize Rothschild and Stiglitz' (\citeyear{RothschildStiglitz1970}) classic analysis of decisionmaking under risk to the multivariate case.

Interesting avenues to explore include determining whether our multivariate majorization approach can be adapted to cases where the principal's objective is non-separable and/or agents' types are multidimensional. We leave these for future research.

\newpage
\appendix
\addtolength{\baselineskip}{-1mm}

\section{Proofs}

\noindent\textbf{Proof of Theorem \ref{iffTtrans}.} For notational convenience, we prove the proposition for the case of a uniform distribution over types for all players. We say that ${\bf y}$ is lexicographically lower than ${\bf x}$, written ${\bf y}\leq_L{\bf x}$, if there is a $k\in N$ such that $y_k>x_k$, then there exists $l\in N$, $l<k$ such that $y_l<x_l$; ${\bf y}$ is strictly lexicographically lower than ${\bf x}$, written ${\bf y}<_L{\bf x}$ if ${\bf y}\leq_L{\bf x}$ and ${\bf x}\neq{\bf y}$. In words, we first sort  type profiles by player, then by type.

 Suppose $g \succ h$. Choose $i\in N$ and ${\bf x}_{-i}$ such that there is $\hat x_i$ and $\tilde x_i$ with $\hat x_i<\tilde x_i$, $h(\hat x_i,{\bf x}_{-i})>g(\hat x_i,{\bf x}_{-i})$, $h(\tilde x_i,{\bf x}_{-i})<g(\tilde x_i,{\bf x}_{-i})$ and where $(\tilde x_i,{\bf x}_{-i})$ is the lexicographically lowest such profile and $(\hat x_i,{\bf x}_{-i})$ is the lexicographically highest such type (lexicographically below $(\tilde x_i,{\bf x}_{-i}))$). Let
\bd
\alpha=\frac{\delta}{g(\tilde x_i, {\bf x}_{-i})-g(\hat x_i,{\bf x}_{-i})}
\ed
where $\delta=\min\big(g(\tilde x_i, {\bf x}_{-i})-h(\tilde x_i,{\bf x}_{-i}),h(\hat x_i, {\bf x}_{-i})-g(\hat x_i,{\bf x}_{-i})\big)$ and define the orthogonal $T$-transform
\bd
T({\bf z}, {\bf y}) =
\left\{
\begin{array}{ll}
\alpha 				& \text{if } {\bf z}_{-i}={\bf y}_{-i} = {\bf x}_{-i}\text{ and } z_i=y_i=\hat x_i \text{ or } z_i=y_i=\tilde x_i \\
1-\alpha 				& \text{if }  {\bf z}_{-i}={\bf y}_{-i} = {\bf x}_{-i}\text{ and } z_i=\hat x_i, y_i=\tilde z_i  \text{ or }  z_i=\tilde x_i, y_i=\hat x_i \\
1	 	& \text{if } {\bf z } = {\bf y} \not\in \{ \hat{\bf x},\tilde{\bf x}\} \\
0		& \text{otherwise}
\end{array}
\right.
\ed
Then
\begin{align*}
(Tg)({\bf z})&
=
\left\{
\begin{array}{ll}
\alpha g(\hat x_i, {\bf x}_{-i})+(1-\alpha)g(\tilde x_i, {\bf x}_{-i})	& \text{if } {\bf z}_{-i}= {\bf x}_{-i}\text{ and } z_i=\hat x_i\\
\alpha g(\tilde x_i, {\bf x}_{-i})+(1-\alpha) g(\hat x_i, {\bf x}_{-i})	& \text{if } {\bf z}_{-i}= {\bf x}_{-i}\text{ and } z_i=\tilde x_i\\
g({\bf z}) &\text{otherwise}
\end{array}
\right.\\
&
=
\left\{
\begin{array}{ll}
 g(\hat x_i,{\bf x}_{-i})	+\delta& \text{if } {\bf z}_{-i}= {\bf x}_{-i}\text{ and } z_i=\hat x_i\\
 g(\tilde x_i, {\bf x}_{-i})-\delta & \text{if } {\bf z}_{-i}= {\bf x}_{-i}\text{ and } z_i=\tilde x_i\\
g({\bf z}) &\text{otherwise}
\end{array}
\right.
\end{align*}
It is clear that $Tg\prec g$. It is also true that $h\prec Tg$. To see this, suppose instead that there is a lower set $X_-\subset X$ such that $\mathbb{E}\big((Tg)({\bf z})|{\bf z} \in X_-\big)>\mathbb{E}\big(h({\bf z})|{\bf z} \in X_-\big)$. Then $(\hat x_i,{\bf x}_{-i})\in X_-$ but $(\tilde x_i,{\bf x}_{-i})\not\in X_-$. By our choice of $i$, $x_{-i}$, $\hat x_i$ and $\tilde x_i$, we may assume
\bd
\{{\bf x}'| {\bf x}'<_L (\tilde x_-,{\bf x}_{-i})\} \subseteq X_-.
\ed
But $(Tg)(\tilde x_i,{\bf x}_{-i})\geq h(\tilde x_i,{\bf x}_{-i})$ so that
\bd
\mathbb{E}\big(Tg({\bf z})|{\bf z} \in X_-\cup\{(\tilde x_i,{\bf x}_{-i})\}\big)>\mathbb{E}\big(h({\bf z})|{\bf z} \in X_-\cup\{(\tilde x_i,{\bf x}_{-i})\}\big)
\ed
and $\mathbb{E}\big(Tg({\bf z})|{\bf z} \in X_-\cup\{(\tilde x_i,{\bf x}_{-i})\}\big)=\mathbb{E}\big(g({\bf z})|{\bf z} \in X_-\cup\{(\tilde x_i,{\bf x}_{-i})\}\big).$ Since $X_-\cup\{(\tilde x_i,{\bf x}_{-i})\}$ is a lower set, this contradicts our assumption that $g\succ h$ and we conclude that $Tg\succ h$.

Finally, for any two functions $\phi$ and $\psi$ on $X$, let $d(\phi,\psi)$ denote the number of type profiles ${\bf x}\in X$ such that $\phi({\bf x})\neq \psi({\bf x})$. Then $d(g,Tg)=d(g,h)-1$. Therefore, repeating this step at most $|X|-1$ times will result in $h$.

Now suppose that $h=Sg$ where $S$ is the product of a sequence of orthogonal $T$-transforms $T^1,\ldots, T^K$ for some finite integer $K$. Suppose $T^1$ mixes between type profiles $(\hat x_i, {\bf x}_{-i})$ and $(\tilde x_i, {\bf x}_{-i})$, with probability $(\alpha,1-\alpha)$ for $\alpha\in [0,1]$ and $\hat x_i<\tilde x_i$. Define
\begin{align*}
g^*({\bf z}) \equiv (T^1g)({\bf z})&
=
\left\{
\begin{array}{ll}
\alpha g(\hat x_i, {\bf x}_{-i})+(1- \alpha)g(\tilde x_i, {\bf x}_{-i})	& \text{if } {\bf z}_{-i}= {\bf x}_{-i}\text{ and } z_i=\hat x_i\\
\alpha g(\tilde x_i, {\bf x}_{-i})+(1- \alpha)g(\hat x_i, {\bf x}_{-i})	& \text{if } {\bf z}_{-i}= {\bf x}_{-i}\text{ and } z_i=\tilde x_i\\
g({\bf z}) &\text{otherwise}
\end{array}
\right..
\end{align*}
Then
\begin{align*}
\mathbb{E}(g^*({\bf x})|x\in X_-)&=
\left\{
\begin{array}{ll}
\mathbb{E}(g({\bf x})|x\in X_-) &\\
\qquad+(1-\alpha)\big(g(\tilde x_i,{\bf x}_{-i})-g(\hat x_i,{\bf x}_{-i})\big)f(\hat x_i,{\bf x}_{-i}) &\text{if } (\hat x_i,{\bf x}_{-i})\in X_-, (\tilde x_i, {\bf x}_{-i})\not\in X_-\\[1mm]
\mathbb{E}(g({\bf x})|x\in X_-) &\text{otherwise}
\end{array}
\right.&\\
&\geq \hspace{6mm}  \mathbb{E}(g({\bf x})|x\in X_-)
\end{align*}
So that $g^*\prec g$. Iterating in this way, we can conclude that $h\prec g$.  $\hfill\blacksquare$

\medskip

\noindent\textbf{Proof of Theorem \ref{form}.} \label{proof: form}The constraint $g\succ\alpha$ in \eqref{eq: ironed alpha} defines a convex polyhedron so the convex program has a unique minimizer, denoted $\overline{\alpha}$.

In order to deal with the majorization constraint $g\succ\alpha$, we parameterize
\begin{equation}\label{gApp}
g(\mathbf{x})\,=\,\alpha(\mathbf{x})-\sum_{i\,\in\,N}\underline{\Delta}_i\lambda_i(\mathbf{x})/f(\mathbf{x})
\end{equation}
where, for  $i\in N$, the $\lambda_i({\bf x})$ are non-negative for ${\bf x}\in X$ with $\lambda_i(\overline{x}_i,{\bf x}_{-i})=0$ for all ${\bf x}_{-i}\in X_{-i}$.
The solutions to \eqref{gApp} must satisfy, for $i\in N$,
\begin{eqnarray}
  \lambda_i^*({\bf x})\overline{\Delta}_ig({\bf x}) &=& 0 \label{compslack}\\
  \overline{\Delta}_ig({\bf x}) &\geq& 0 \label{posder}
\end{eqnarray}
which imply $\lambda_i^*({\bf x})\overline{\Delta}_i\phi'(g({\bf x}))=0$ and $\overline{\Delta}_i\phi'(g({\bf x}))\geq 0$ for any convex $\phi(x)$ with $\phi(0)=\phi'(0)=0$.

Suppose, in contradiction, that $\overline{\alpha}$ is not non-decreasing, i.e. $\overline{\Delta}_i\overline{\alpha}({\bf x})<0$ for some $\mathbf{x}\in\mathbf{X}$,
$i\in N$. Then lowering $\overline{\alpha}({\bf x})$ slightly while raising $\overline{\alpha}(x_i^+,{\bf x}_{-i})$ by the same amount lowers the objective in \eqref{eq: ironed alpha}, a contradiction.

Suppose, in contradiction, that $\overline{\alpha}$ is not minimal, i.e. there exists a non-decreasing $g({\bf x})$ such that $\overline{\alpha}\succ g\succ\alpha$. By Theorem \ref{iffTtrans} there exists a doubly stochastic operator $T({\bf x},{\bf y})$ such that $g({\bf x})=\sum_{{\bf y}\in X}T({\bf x},{\bf y})\overline{\alpha}({\bf y})$.  But, convexity of the objective in \eqref{eq: ironed alpha} then implies that $g$ yields a lower value than $\overline{\alpha}$, a contradiction.

Next, suppose, in contradiction, that the level sets of $\overline{\alpha}({\bf x})$ do not form an ultramodular partition, i.e. $\overline{\alpha}(\mathbf{x}_1)=\overline{\alpha}(\mathbf{x}_2)=\overline{\alpha}^*$
for some $\mathbf{x}_2\geq\mathbf{x}_1$ but $\overline{\alpha}(\mathbf{x})\neq\overline{\alpha}^*$ for some $\mathbf{x}_1\leq\mathbf{x}\leq\mathbf{x}_2$. If either $\overline{\alpha}(\mathbf{x})<\overline{\alpha}^*=\overline{\alpha}(\mathbf{x}_1)$ or
$\overline{\alpha}(\mathbf{x})>\overline{\alpha}^*=\overline{\alpha}(\mathbf{x}_2)$ then $\overline{\alpha}$ is not non-decreasing.

Finally we show that $q^*$ does indeed solve the original problem \ref{problem: original problem}. Note that any non-decreasing $q(\mathbf{x})$ can be written as
$q(\mathbf{x})=\beta-\sum_{\mathbf{X}_-\subset\,\mathbf{X}}\beta_{\mathbf{X}_-}I_{\mathbf{X}_-}(\mathbf{x})$ where the sum is over all lower sets of $\mathbf{X}$ that are strict subsets of $\mathbf{X}$, the $\beta$s are scalars with $\beta\in\mathbb{R}$ and $\beta_{\mathbf{X}_-}\geq 0$, and the $I_{\mathbf{X}_-}$ are indicator functions with $I_{\mathbf{X}_-}(\mathbf{x})=1$ if $\mathbf{x}\in\mathbf{X}_-$ and $I_{\mathbf{X}_-}(\mathbf{x})=0$ otherwise.
The intuition behind the majorization conditions in Definition \ref{multiVariate} is that they ensure, for any $\overline{\alpha}\succ\alpha$, that $\mathbb{E}[q(\mathbf{x})(\overline{\alpha}(\mathbf{x})-\alpha(\mathbf{x}))]\geq 0$ and, hence,
$\mathbb{E}[q(\mathbf{x})\overline{\alpha}(\mathbf{x})-C(q(\mathbf{x}))]\geq\mathbb{E}[q(\mathbf{x})\alpha(\mathbf{x})-C(q(\mathbf{x}))]$. And since we wish the latter to hold with equality, the sought after $\overline{\alpha}$ must be a minimal element.

Next consider a partition $\mathscr{P}$ of $\mathbf{X}$ and denote the cell of $\mathscr{P}$ that $\mathbf{x}$ belongs to by $P(\mathbf{x})$, which may be a singleton or contain multiple types. Suppose for $\mathbf{y}\in P(\mathbf{x})$ we replace $\alpha(\mathbf{y})$ with $\overline{\alpha}(\mathbf{y})=\mathbb{E}[\alpha(\mathbf{y})|\,\mathbf{y}\in P(\mathbf{x})]$ and that $\overline{\alpha}$ is non-decreasing, majorizes $\alpha$, and is minimal.
The $q^*(\mathbf{x})$ that maximizes \eqref{problem: relaxed problem}
is then constant on any of the partition cells. Hence,
\begin{eqnarray*}
   \mathbb{E}[q^*(\mathbf{x})(\overline{\alpha}(\mathbf{x})-\alpha(\mathbf{x}))] &=&  \underset{P_i\,\in\,\mathscr{P}}{\mathbb{E}}\,\bigl[\mathbb{E}[q^*(\mathbf{x})(\overline{\alpha}(\mathbf{x})-\alpha(\mathbf{x}))\,|\,\mathbf{x}\,\in\,P_i]\bigr]\\[1mm]
   &=& \underset{P_i\,\in\,\mathscr{P}}{\mathbb{E}}\,\bigl[q^*(\mathbf{x})\,\mathbb{E}[\,\overline{\alpha}(\mathbf{x})-\alpha(\mathbf{x})\,|\,\mathbf{x}\,\in\,P_i]\bigr]\\[1mm]
   &=& \,\,\,\,\,0
\end{eqnarray*}
i.e. there is ``no gap:'' $\mathbb{E}[q^*(\mathbf{x})\overline{\alpha}(\mathbf{x})-C(q^*(\mathbf{x}))]=\mathbb{E}[q^*(\mathbf{x})\alpha(\mathbf{x})-C(q^*(\mathbf{x}))]$. Moreover, for any non-decreasing $q(\mathbf{x})$ we have
\begin{eqnarray*}
  \mathbb{E}[q(\mathbf{x})\alpha(\mathbf{x})-C(q(\mathbf{x}))] &\leq& \mathbb{E}[q(\mathbf{x})\overline{\alpha}(\mathbf{x})-C(q(\mathbf{x}))] \\[1mm]
  &\leq& \mathbb{E}[q^*(\mathbf{x})\overline{\alpha}(\mathbf{x})-C(q^*(\mathbf{x}))] \\[1mm]
  &=& \mathbb{E}[q^*(\mathbf{x})\alpha(\mathbf{x})-C(q^*(\mathbf{x}))]
\end{eqnarray*}
where the first inequality follows from $\overline{\alpha}\succ\alpha$, the second inequality from optimality of $q^*$ for $\overline{\alpha}$, and the final equality from the ``no gap'' result. Together they show $q^*$ solves problem \ref{problem: original problem}.

 $\hfill\blacksquare$

\medskip

\noindent\textbf{Proof of Theorem \ref{thm: continuous without access}.} By Theorem \ref{form}, $\overline{\alpha}_n$ is non-decreasing. Let $D$ be the vertices of all dyadic cubes. Because $\alpha$ is bounded, it follows that $\overline{\alpha}_n$ is uniformly bounded and there exists a subsequence $\overline{\alpha}_{n_k}$ such that pointwise convergents on $D$ to a function $\phi$, i.e., $\phi(\mathbf{x})=\lim \limits_{k \to \infty} \overline{\alpha}_{n_k}(\mathbf{x}), \forall \mathbf{x}\in D$. Extend $\phi$ to $\overline{\alpha}(\mathbf{x})$ on entire $X$:
\bd
\overline{\alpha}(\mathbf{x}) = \left\{\begin{array}{ll}
	\phi(\bf x) & \text{\em if } \mathbf{x}\,\in\,D\\
	\sup_{\mathbf{t}\leq\mathbf{x},\mathbf{t}\in D} \phi(\mathbf{t}) & \text{otherwise}
\end{array}
\right.
\ed
By definition, $\overline{\alpha}$ is non-decreasing. Let the set of discontinuous points of $\overline{\alpha}$ be $\mathcal{D}(\overline{\alpha})$, and note that $\mathcal{D}(\overline{\alpha})$ is countable, see \cite{BBS2004}. When $\mathbf{x}\notin \mathcal{D}(\overline{\alpha})$, then $\overline{\alpha}(\mathbf{x})=\lim \limits_{k \to \infty} \overline{\alpha}_{n_k}(\mathbf{x})$. To see this, by density of $D$ and continuity of $\overline{\alpha}$ at $\mathbf{x}$, for $\forall \epsilon>0$, there exist $\mathbf{p},\mathbf{q} \in D$ such that $\mathbf{p}<\mathbf{x}<\mathbf{q}$ and $\overline{\alpha}(\mathbf{q})-\overline{\alpha}(\mathbf{p})<\frac{\epsilon}{2}$. Due to pointwise convergence of $\overline{\alpha}_{n_k}$ at $D$, if $k$ is sufficiently large, then $\overline{\alpha}_{n_k}(\mathbf{q})\in(\overline{\alpha}(\mathbf{q})-\frac{\epsilon}{2},\overline{\alpha}(\mathbf{q})+\frac{\epsilon}{2})$ and $\overline{\alpha}_{n_k}(\mathbf{p})\in(\overline{\alpha}(\mathbf{p})-\frac{\epsilon}{2},\overline{\alpha}(\mathbf{p})+\frac{\epsilon}{2})$. Also by the monotonicity of $\overline{\alpha}$, $\overline{\alpha}(\mathbf{p})\leq\overline{\alpha}(\mathbf{x})\leq\overline{\alpha}(\mathbf{q})$. Since $\overline{\alpha}_{n_k}$ are non-decreasing,
\begin{align*}
\overline{\alpha}_{n_k}(\mathbf{x})\leq \overline{\alpha}_{n_k}(\mathbf{q})<\overline{\alpha}(\mathbf{q})+\frac{\epsilon}{2}<\overline{\alpha}(\mathbf{p})+\epsilon<\overline{\alpha}(\mathbf{x})+\epsilon \\
\overline{\alpha}_{n_k}(\mathbf{x}) \geq \overline{\alpha}_{n_k}(\mathbf{p})>\overline{\alpha}(\mathbf{p})-\frac{\epsilon}{2}>\overline{\alpha}(\mathbf{q})-\epsilon>\overline{\alpha}(\mathbf{x})-\epsilon
\end{align*}
Let $\epsilon \to 0$, then $\overline{\alpha}(\mathbf{x})=\lim \limits_{k \to \infty} \overline{\alpha}_{n_k}(\mathbf{x}),\mathbf{x}\notin \mathcal{D}(\overline{\alpha})$.

Let $\Omega_k$ be any lower set of $\mathscr{D}_k$, choose some $n_k > k$, it follows that
\bd
\int_{\Omega_k} \overline{\alpha}_{n_k}\mathrm{d}F\,\leq\,\int_{\Omega_k} \alpha_{n_k}\mathrm{d}F\,=\,\int_{\Omega_k} \alpha\mathrm{d}F
\ed
with equality if $\Omega_k = X$. First let $n_k \to \infty$ in above inequality, by dominated convergence theorem, it follows that
$\int_{\Omega_k} \overline{\alpha}\mathrm{d}F\leq\int_{\Omega_k} \alpha\mathrm{d}F$. Due to continuity from below of $F$, i.e. $\Omega_k \uparrow \Omega \implies F(\Omega_k)\uparrow F(\Omega)$, then let $k \to \infty$, we obtain $\int_\Omega \overline{\alpha}\mathrm{d}F\,\leq\,\int_\Omega \alpha\mathrm{d}F$ for any lower set $\Omega \in \mathscr{B}$. Thus, $\overline{\alpha}\succ\alpha$.

Next we show Problem \eqref{problem: original problem} and \eqref{problem: relaxed problem} are equivalent under $\overline{\alpha}$. We denote the objective function of Problem \eqref{problem: original problem} and \eqref{problem: relaxed problem} by $\Pi(q,\alpha)\,=\,q(\mathbf{x})\alpha(\mathbf{x})-C(q)$. As $\alpha_n \xrightarrow{\text{a.e.}}\alpha$, by bounded convergence theorem, $\Pi(q,\alpha_n)\xrightarrow{\text{a.e.}}\Pi(q,\alpha)$. Note that $\Pi(q_n^*,\alpha_n)\,=\,\Pi(q_n^*,\overline{\alpha}_n)$, $q_n^*\,=\,C'^{-1}(\max\{\overline{\alpha}_n,0\})$ by Theorem \ref{form}, and $q^*_n \xrightarrow{\text{a.e.}}q^*\,=\,C'^{-1}(\max\{\overline{\alpha},0\})$, it follows that $\Pi(q^*_{n_k},\alpha_{n_k})\,=\,\Pi(q^*_{n_k},\overline{\alpha}_{n_k})$ and let $k\to\infty$ on both sides, we get $\Pi(q^*,\alpha)\,=\,\Pi(q^*,\overline{\alpha})$.$\hfill\blacksquare$

We can obtain stronger results with more mild assumptions on $\alpha$.
\begin{corollary}\label{cor: a.e. convergent}
	If $\alpha$ is a continuous function of bounded variation, i.e.,
	\bd
	\exists K\,>\,0, \ \sup\limits_{\cup d^N_n = X} \Bigl\{\sum_{d^N_n} |\alpha(\mathbf{x})-\alpha(\mathbf{y})| \, \big| \, \mathbf{x,y}\,\in\,d^N_n\Bigr\}\,<\,K
	\ed
	 $\overline{\alpha}_n$ converge a.e. to $\overline{\alpha}$.
\end{corollary}
\noindent\textbf{Proof of Corollary \ref{cor: a.e. convergent}.} $|\overline{\alpha}_n(\mathbf{x})-\overline{\alpha}(\mathbf{x})|\,\leq\,|\overline{\alpha}_n(\mathbf{x})-\overline{\alpha}_{n_k}(\mathbf{x})|+|\overline{\alpha}_{n_k}(\mathbf{x})-\overline{\alpha}(\mathbf{x})|$. By Theorem \ref{thm: continuous without access}, the second term on RHS tend to $0$ a.e., we show the first term is a.e. a Cauchy sequence. The value jump of the ironed part of both $\overline{\alpha}_n$ and $\overline{\alpha}_{n_k}$ tend to $0$ and $|\overline{\alpha}_n(\mathbf{x})-\overline{\alpha}_{n_k}(\mathbf{x})|\to0$ on the intersections of non-flat parts of $\overline{\alpha}_n$ and $\overline{\alpha}_{n_k}$ if $n,n_k$ large enough. Also note that the flat parts become stable because $F$ is absolute continuous. In addition, the jump points are countable and the result follows. $\hfill\blacksquare$

\medskip

\noindent\textbf{Proof of Theorem \ref{thm: iron with access rights}.}
Let us write the principal's problem as a saddle point problem. The constraint that $\eta_i({\bf x})q({\bf x})$ is non-decreasing in $x_i$ for all $i\in N$ can be dealt with by adding $\sum_{i\in N,{\bf x}\in X}\lambda_i({\bf x})\eta_i({\bf x})\overline{\Delta}q_i({\bf x})$ where the $\lambda_i({\bf x})$ are non-negative for all ${\bf x}\in X$ with $\lambda_i(\overline{x}_i,{\bf x}_{-i})=0$ for all $i\in N$. This term can rewritten to yield the following saddle-point problem
\begin{equation} \label{saddleMulti}
\Pi\,=\,\min_{\lambda\,:\,X\,\rightarrow\,\mathbb{R}_+}\,
\max_{(q,{\bm \eta})\,:\,X\,\rightarrow\,\mathbb{R}_{\geq 0}\times [0,1]^{n}}
\mathbb{E}\Bigl[q({\bf x})\sum_{i\,\in\,N}\eta_i({\bf x})(\alpha_i({\bf x})-\underline{\Delta}_i\lambda_i({\bf x})/f({\bf x}))-C(q({\bf x}))\Bigr]
\end{equation}

We construct the partition used in the proposition and prove it is ortho-convex. First note that for any player, or any dimension, $i$, while fixing $\mathbf{x}_{-i}$, $i$'s type space is partitioned into intervals $P$ such that (i) $\lambda_i(x_i,{\bf x}_{-i}) = 0$ for $x_i<\min\{x_i'|(x_i',{\bf x}_{-i})\in P\}$; (ii) $\lambda_i(x_i,{\bf x}_{-i}) = 0$ for $x=\max\{x_i'|(x_i, {\bf x}_{-i})\in P\}$; and (iii) if $|P|>1$ then $\lambda_i(x_i,{\bf x}_{-i})>0$ for $ \min\{P\}\leq x<\max\{P\}$. Denote this partition by $\mathscr{P}_i$.


The following algorithm constructs the ironed or flattened subset of $X$ containing $P\in \mathscr{P}_i$. Let $P^1 = P$. By the Kuhn-Tucker conditions associated with problem \eqref{saddleMulti}, any feasible mechanisms requires that $\eta_i(x_i,\mathbf{x}_{-i})q(x_i,\mathbf{x}_{-i})$ be constant in $i$ for all $\mathbf{x}\in P$. Moreover, $\mathbb{E}(\underline\Delta_i\lambda_i(\mathbf{x})|\mathbf{x}\in P^1) = 0$.

Given $P^{k-1}$, define
\[
P^k = P^{k-1} \bigcup_{j\in N} \big\{\hat P \in \mathscr P_j\big| \hat P\not\subseteq P^{k-1}, \hat P\cap P^{k-1}\neq \emptyset\big\}.
\]
The set $P^k$ adds to $P^{k-1}$ all intervals with positive multipliers in all dimensions that intersect with $P^{k-1}$ (but are not already contained in $P^{k-1}$).

The solution to \eqref{ironedMultiWith} must be such that $\eta_j(\cdot)q(\cdot)$ is constant on $P^k$. To see this, suppose $\eta_j(\cdot)q(\cdot)=c$ is constant on $P^{k-1}$ and that there is some $j\in N$ and $\hat P\in\mathscr{P}_j$ such that $\hat P\not\subseteq P^{k-1}$ and $\hat P\cap P^{k-1}\neq \emptyset$ -- i.e. an interval in dimension $j$ that intersects with and is not inside $P^{k-1}$. Then $\eta_j(\cdot)q(\cdot)=c$ on $\hat P\cap P^{k-1}$ by assumption. Using the Kuhn-Tucker conditions associated with problem \eqref{saddleMulti} in dimension $j$, this extends to the entire set $\hat P$.

Each iterative set is ortho-convex, after possibly adding some knife-edge type profiles where the monotonicity constraint just binds: take $(x_j,\mathbf{x}_{-j})\in P^k $ and $(x'_j,\mathbf{x}_{-j})\in P^{k}$ with $x_j'>x_j$. Since any incentive compatible mechanism requires that $\eta_j(\cdot)q_j(\cdot)$ be non-decreasing in dimension $j$, $\eta_j(x'_j,\mathbf{x}_{-j})q(x'_j,\mathbf{x}_{-j})\geq \eta_j(\hat x_j,\mathbf{x}_{-j})q(\hat x_j,\mathbf{x}_{-j})\geq \eta_j( x_j,\mathbf{x}_{-j})q( x_j,\mathbf{x}_{-j})$ for any $\hat x_j \in (x_j, x_j')$. But $\eta_j(x'_j,\mathbf{x}_{-j})q(x'_j,\mathbf{x}_{-j})= \eta_j( x_j,\mathbf{x}_{-j})q( x_j,\mathbf{x}_{-j})$ so $\eta_i(\hat x_j,\mathbf{x}_{-j})q(\hat x_j,\mathbf{x}_{-j})= \eta_i( x_j,\mathbf{x}_{-j})q( x_j,\mathbf{x}_{-j})$ for any $\hat x_j \in (x_j, x_j')$. If the interval $[x_j,x_j']\times \{\mathbf{x}_{-j}\}\not\subseteq P^k$, add it to the set before moving to the next step.

The algorithm ends when the set $\big\{\hat P \in \mathscr{P}_j\big| \hat P\not\subseteq P^{k}, \hat P\cap P^{k}\neq \emptyset\big\}$ is empty for all players (i.e. when no intervals not inside the iterative set intersect with the iterative set). Call the resulting set $P$. Then $P$ is ortho-covex and $\mathbb{E}(\sum_{j\in N}\underline\Delta_j\lambda_j(\mathbf{x})|\mathbf{x}\in P) = 0$. To see the latter, suppose $k$ was the final step in the algorithm and that $\big\{\hat P \in \mathscr{P}_l\big| \hat P\not\subseteq P^{k}, \hat P\cap P^{k}\neq \emptyset\big\}$ is empty for all players $l\in N$ but there exists $j\in N$ such that $\mathbb{E}(\underline\Delta_j\lambda_j(\mathbf{x})|\mathbf{x}\in P^k) \neq 0$. Then, there must be $\mathbf{x}\in P^k\cap\hat P$ for some $\hat P \in \mathscr{P}_j$ by definition of $\mathscr{P}_j$. But, since $\mathbb{E}(\underline\Delta_j\lambda_j(\mathbf{x})|\mathbf{x}\in  P') = 0$ for any $P'\in \mathscr{P}_j$, it must be that $\hat P\not\subseteq P^k$. Therefore $\hat P\cap P \neq \emptyset$ and $\hat P\not\subseteq P$, contradicting the assumption that $\big\{\hat P \in \mathscr{P}_j\big| \hat P\not\subseteq P^{k-1}, \hat P\cap P^{k-1}\neq \emptyset\big\}$ is empty.

Repeating this algorithm for all $P\in \mathscr{P}_i$ and all $i\in N$ results in a partition of the type space; denote this partition by $\mathscr{P}$. Define $P(\mathbf{x}) = \{ \mathbf{y}\in P | \mathbf{x} \in P, P\in \mathscr{P}\}$.

Next, we show the saddle-point problem for restricted rights has the strong duality property under the choice of $q^*(\mathbf{x})$ and $\eta_i^*(\mathbf{x})$. From above, we have that for any cell $P_i\in\mathscr{P}_i$, $P_i$ is contained in one cell of $\mathscr{P}$, and, for fixed ${\bf x}_{-i}$, $q^*\eta_i^*$ is constant on $P_i$. Because on each ironed interval $P_i$, $\mathop{\mathbb{E}} [\underline{\Delta}_i \lambda_i^*(x_i,\mathbf{x_{-i}})|(x_i,\mathbf{x_{-i}})\in P_i]=0$, we have
\[
\mathop{\mathbb{E}} [\widetilde{\alpha}_i(x_i,\mathbf{x_{-i}})|(x_i,\mathbf{x_{-i}})\in P_i]=\mathop{\mathbb{E}} [\alpha_i(x_i,\mathbf{x_{-i}})|(x_i,\mathbf{x_{-i}})\in P_i].
\]
We calculate the value gap between the saddle-point problem and the primal problem under $(q^*,\bm{\eta}^*,\bm{\lambda}^*)$:
\begin{align*}
 \pi(\widetilde{{\bm\alpha}},q^*, &{\bm \eta}^*)-\pi(\widetilde{{\bm\alpha}},q^*, {\bm \eta}^*)\\
    =&\mathop{\mathbb{E}} \left[q^*(\mathbf{x})\sum_i \widetilde{\alpha}_i(\mathbf{x})\eta_i^*(\mathbf{x})\right]-\mathop{\mathbb{E}} \left[q^*(\mathbf{x})\sum_i \alpha_i(\mathbf{x})\eta_i^*(\mathbf{x})\right]\\
    =&\mathop{\mathbb{E}} \left[\sum_i q^*(\mathbf{x})\eta_i^*(\mathbf{x})(\widetilde{\alpha}_i(\mathbf{x})-\alpha_i(\mathbf{x}))\right]\\
    =&\sum_i \mathop{\mathbb{E}}\left[ q^*(\mathbf{x})\eta_i^*(\mathbf{x})(\widetilde{\alpha}_i(\mathbf{x})-\alpha_i(\mathbf{x}))\right]\\
    =&\sum_i \mathop{\mathbb{E}}\left[\mathop{\mathbb{E}}[q^*(x_i,\mathbf{x}_{-i})\eta_i^*(x_i,\mathbf{x_{-i}})(\widetilde{\alpha}_i(x_i,\mathbf{x_{-i}})-\alpha_i(x_i,\mathbf{x}_{-i}))|(x_i,\mathbf{x}_{-i})\in P_i]\right]\\
    =&\sum_i \mathop{\mathbb{E}}\left[q^*(x_i,\mathbf{x}_{-i})\eta_i^*(x_i,\mathbf{x}_{-i})\mathop{\mathbb{E}}[\widetilde{\alpha}_i(x_i,\mathbf{x}_{-i})-\alpha_i(x_i,\mathbf{x}_{-i})|(x_i,\mathbf{x}_{-i})\in P_i]\right]=0
\end{align*}
We thus have
\begin{displaymath}
  \pi({\bm \alpha},q, {\bm \eta})\,\leq\,\pi(\widetilde{{\bm\alpha}},q, {\bm \eta})\,\leq\,\pi(\widetilde{{\bm\alpha}},q^*, {\bm \eta}^*)\,=\,\pi({\bm \alpha},q^*, {\bm \eta}^*)
\end{displaymath}
where the first inequality follows because, for every $i\in N$, $\widetilde{\alpha}_i\succ_i \alpha_i$ and $q(\cdot)\eta_i(\cdot)$ is non-decreasing implies $\mathbb{E}[q({\bf x})\eta_i({\bf x})(\alpha_i({\bf x})-\widetilde{\alpha}_i({\bf x}))]\leq 0$, the second inequality follows from optimality of $(q^*,{\bm \eta}^*)$ for $\widetilde{{\bm\alpha}}$, and the final equality follows from the zero value gap. Hence, $(q^*,{\bm \eta}^*)$ is optimal for ${\bm \alpha}$. Therefore, strong duality holds for the saddle-point problem with restricted access rights, $(q^*,\bm{\eta}^*,\bm{\lambda}^*)$ is a saddle-point for the Lagrangian, and $(q^*,\bm{\eta}^*)$ solves the primal problem.$\hfill\blacksquare$

\medskip

\begin{lemma}\label{multiplicity}
If there exist two solutions, $\widetilde{\bm \alpha}$ and $\widetilde{{\bm\alpha}}'$, to \eqref{ironedMultiWith} then the associated optimal mechanisms $(q^*,\bm{\eta}^*,\bf{t}^*)$ and  $(\hat q^*,\hat{\bm{\eta}}^*,\hat{\bf{t}}^*)$ are identical.
\end{lemma}

\noindent\textbf{Proof of Lemma \ref{multiplicity}.} Since both $\widetilde{{\bm\alpha}}$ and $\widetilde{{\bm\alpha}}'$ are assumed to achieve the maximum in problem \eqref{ironedMultiWith} and since $\sum_{i\,\in\,N}\max\{0,g_i({\bf x})\}^2$ is strictly convex over the positive range of $g_i$, we can assume that $\widetilde{\alpha}_i({\bf x})\leq 0$ and $\widetilde{\alpha}'_i({\bf x})\leq 0$ whenever $\widetilde{\alpha}_i({\bf x})\neq \widetilde{\alpha}'_i({\bf x})$. Otherwise, for the convex combination $\widetilde{{\bm\alpha}}''=\frac{1}{2}( \widetilde{{\bm\alpha}}+\widetilde{{\bm\alpha}}')$,
\bd
\mathbb{E}\Bigl[\bigl(\,\sum_{i\,\in\,N}\max(0,\widetilde{\alpha}_i(\cdot))\bigr)^2\Bigr]>\mathbb{E}\Bigl[\bigl(\,\sum_{i\,\in\,N}\max(0,\widetilde{\alpha}''_i(\cdot))\bigr)^2\Bigr]
\ed
and $\widetilde{\alpha}''_i \succ M\!R_i$. Moreover, if $\widetilde{\alpha}_i({\bf x})< 0$ and $\widetilde{\alpha}'_i({\bf x})< 0$ whenever $\widetilde{\alpha}_i({\bf x})\neq \widetilde{\alpha}'_i({\bf x})$, then $\eta_i({\bf x})=\hat \eta_i({\bf x})=0$.

Suppose then that there exists $(x_i,{\bf x}_{-i})\in X$ such that $\widetilde{\alpha}'_i(x_i,{\bf x}_{-i})<\widetilde{\alpha}_i(x_i,{\bf x}_{-i})\leq 0$ and without loss of generality suppose it is the largest type with $\widetilde{\alpha}'_i(x_i,{\bf x}_{-i})\neq \widetilde{\alpha}_i(x_i,{\bf x}_{-i})$. Then $\hat \eta_i^*({\bf x})=0$ and $\hat q^*(x'_i,{\bf x}_{-i})\hat \eta_i^*(x'_i,{\bf x}_{-i})=0$ for any $x'_i<x_i$. But this implies that $\widetilde{\alpha}'_i(x_i,{\bf x}_{-i})\leq 0$ for all $x_i'<x_i$ (since otherwise Theorem \ref{thm: iron with access rights} would require that $\hat\eta^*_i(x'_i,{\bf x}_{-i})>0$). Similarly, if $\widetilde{\alpha}_i(x_i,{\bf x}_{-i})< 0$ then  $q^*(x'_i,{\bf x}_{-i}) \eta_i^*(x'_i,{\bf x}_{-i})$ and we are done.

Suppose instead that $\widetilde{\alpha}_i(x_i,{\bf x}_{-i})=0$. There exists $x_i'$ such that $\widetilde{\alpha}(x_i',{\bf x}_{-i})<\widetilde{\alpha}'_i(x_i',{\bf x}_{-i})\leq 0$ since $\widetilde{\alpha}_i$ and $\widetilde{\alpha}'_i$ must have the same expectation below $x_i$. Since $x_i$ is the largest type for which the two solutions differ, $\lambda_i(x_i,{\bf x}_{-i})=\hat \lambda_i(x_i,{\bf x}_{-i})$. Further, since $\widetilde{\alpha}'_i(x_i,{\bf x}_{-i})<\widetilde{\alpha}_i(x_i,{\bf x}_{-i})$, $\lambda_i(x_i^-,{\bf x}_{-i})>\hat\lambda_i(x_i^-,{\bf x}_{-i})\geq 0$; that is, the monotonicity constraints binds between for $x_i^-$ and $x_i$. Furthermore, either $\lambda_i((x_i^-)^-,{\bf x}_{-i})>\hat\lambda_i((x_i^-)^-,{\bf x}_{-i})\geq 0$ or $\widetilde{\alpha}_i(x_i^-,{\bf x}_{-i})<\widetilde{\alpha}'_i(x_i^-,{\bf x}_{-i})\leq 0$. In words, either the $\widetilde{\alpha}_i$ is negative for the type just below $x_i$ or the monotonicity constraints bind between for $(x_i^-)^-$, $x_i^-$ and $x_i$. Iterating in this way, we can conclude that $\lambda_i(\tilde x_i,{\bf x}_{-i})>0$ for all $\tilde x_i\in [x_i',x_i^-]$ which implies that $\eta_i^*(\tilde x_i,{\bf x}_{-i})q^*(\tilde x_i,{\bf x}_{-i})$ is constant over this interval. Since $\widetilde{\alpha}_i(x_i^-,{\bf x}_{-i})<0$, Proposition \ref{thm: iron with access rights} requires $\eta_i^*(\tilde x_i,{\bf x}_{-i})q^*(\tilde x_i,{\bf x}_{-i})=0$.$\hfill\blacksquare$

\medskip

\noindent\textbf{Proof of Theorem \ref{sweep2DS}.} We have
\begin{displaymath}
  F(x)\,=\,\int_0^1\int_0^xt(z,y)dzdG(y)\,=\,\int_0^xt(z,1)dz-\int_0^1\int_0^x\partial_yt(z,y)G(y)dzdy
\end{displaymath}
The first term on the far right side is equal to zero when $x<1$ and it is equal to $G(1)=1$ when $x=1$. Let $\Theta(x)=1$ $x\geq 0$ and $\Theta(x)=0$ for $x<0$ with $\Theta'(x)=\delta(x)$. Then
\begin{displaymath}
  F(x)\,=\,\int_0^1T(x,y)G(y)dy
\end{displaymath}
where we defined
\begin{displaymath}
  T(x,y)\,=\,\Theta(x-1)\delta(1-y)-\int_0^x\partial_yt(z,y)dz
\end{displaymath}
We first show that $T(x,y)$ is doubly stochastic. We have
\begin{eqnarray*}
  \int_0^1T(x,y)dx &=& -\partial_y\int_0^1\int_0^xt(z,y)dzdx \\
   &=& -\partial_y\int_0^1(1-z)t(z,y)dz \\
   &=& -\partial_y(1-y)\,\,=\,\,1
\end{eqnarray*}
for all $y\in[0,1]$, and
\begin{displaymath}
  \int_0^1T(x,y)dy\,=\,\int_0^x(t(z,0)-t(z,1))dz+\Theta(x-1)\,=\,1
\end{displaymath}
for all $x\in[0,1]$. Conversely, if $F(x)=\int_0^1T(x,y)G(y)dy$ then
\begin{eqnarray*}
  f(x) &=& -\int_0^1\partial_yt(x,y)G(y)dy+\int_0^1\delta(x-1)\delta(1-y)G(y)dy \\
   &=& \int_0^1t(x,y)g(y)dy-t(x,1)+\delta(x-1) \\
   &=& \int_0^1t(x,y)g(y)dy
\end{eqnarray*}
since $t(x,1)=\delta(1-x)=\delta(x-1)$.$\hfill\blacksquare$

\medskip

\noindent\textbf{Proof of Theorem \ref{theorem:SOSD}.} Using partial integration, we have
\begin{eqnarray*}
  \int_{[0,1]^N}u({\bf x})g({\bf x})d{\bf x} &=& (-1)^{N+1}\int_{[0,1]^N}u({\bf x})D_{I_N}\overline{G}({\bf x})d{\bf x}\nonumber\\[1mm]
  &=& u(1,\ldots,1)-\sum_{S\,\in\,2^{I_N}}\int_{[0,1]^{|S|}}D_Su({\bf x}_S)\overline{G}_S({\bf x}_S)d{\bf x}_S \\[1mm]
  &\geq& u(1,\ldots,1)-\sum_{S\,\in\,2^{I_N}}\int_{[0,1]^{|S|}}D_Su({\bf x}_S)\overline{F}_S({\bf x}_S)d{\bf x}_S\,\,=\,\,\int_{[0,1]^N}u({\bf x})f({\bf x})d{\bf x}
\end{eqnarray*}
The inequality in the third line follows from two observations. First, since $-D_Su({\bf x}_S)$ is non-decreasing in each argument it can be written as a linear combination, with non-negative coefficients, of indicator functions of upper sets $U_S$, which are upper sets in the hypercube's boundary with the $i$-th coordinate zero if $i\not\in S$. Second, \eqref{Xplus2} implies $\int_{L}\overline{G}({\bf x})d\mathbf{x}\leq\int_{L}\overline{F}({\bf x})d\mathbf{x}$ for any lower set $L\subseteq[0,1]^N$. In particular, for lower sets $L_S$ in the hypercube's boundary where the $i$-th coordinate is zero if $i\not\in S$. And since $\int_{[0,1]^{|S|}}\overline{G}_S({\bf x}_S)d{\bf x}_S=\int_{[0,1]^{|S|}}\overline{F}_S({\bf x}_S)d{\bf x}_S$ we have $\int_{U_S}\overline{G}({\bf x})d\mathbf{x}\geq\int_{U_S}\overline{F}({\bf x})d\mathbf{x}$ for any
upper set $U_S$ in the hypercube's boundary with the $i$-th coordinate zero if $i\not\in S$. Combining these two observations establishes the inequality.

Conversely, suppose $\int_{[0,1]^N}u({\bf x})g({\bf x})d{\bf x}\geq\int_{[0,1]^N}u({\bf x})f({\bf x})d{\bf x}$ for all ortho-concave $u$. Take
\begin{displaymath}
  u(\mathbf{x})\,=\,\int_0^{x_1}\cdots\int_0^{x_N}I_{\mathbf{X}_-}(\mathbf{y})d\mathbf{y}
\end{displaymath}
where $I_{\mathbf{X}_-}(\mathbf{y})$ is the indicator function for the lower set $\mathbf{X}_-$ as before. Then
\begin{displaymath}
  \int_{[0,1]^N}\!\!\!u({\bf x})g({\bf x})d{\bf x}\,=\,(-1)^{N+1}\int_{[0,1]^N}\!\!\!u(\mathbf{x})D_{I_N}\overline{G}(\mathbf{x})d\mathbf{x}\,=\,
  -\int_{[0,1]^N}\!\!\!D_{I_N}u(\mathbf{x})\overline{G}(\mathbf{x})d\mathbf{x}\,=\,-\int_{\mathbf{X}_-}\!\!\overline{G}(\mathbf{x})d\mathbf{x}
\end{displaymath}
and, similarly, $\int_{[0,1]^N}u({\bf x})f({\bf x})d{\bf x}=-\int_{\mathbf{X}_-}\overline{F}(\mathbf{x})d\mathbf{x}$. Hence, $\int_{\mathbf{X}_-}\overline{G}(\mathbf{x})d\mathbf{x}\leq\int_{\mathbf{X}_-}\overline{F}(\mathbf{x})d\mathbf{x}$ for any lower set $\mathbf{X}_-$.
$\hfill\blacksquare$

\newpage
\addtolength{\baselineskip}{0.45mm}

\bibliographystyle{klunamed}
\bibliography{library}

\end{document}